# Nematic superconducting state in iron pnictide superconductors

Jun Li,[1,2,3†] Paulo J. Pereira,[3,4†] Jie Yuan,[5,1,2†] Yang-Yang Lv,[1] Mei-Ping Jiang,[1] Dachuan Lu,[1] Zi-Quan Lin,[6] Yong-Jie Liu,[6] Jun-Feng Wang,[6] Liang Li,[6] Xiaoxing Ke,[7] Gustaaf Van Tendeloo,[7] Meng-Yue Li,[1,2] Hai-Luke Feng,[2] Takeshi Hatano,[2] Hua-Bing Wang,[1,2*] Pei-Heng Wu,[1,8] Kazunari Yamaura,[2,9*] Eiji Takayama-Muromachi,[2,9] Johan Vanacken,[3] Liviu F. Chibotaru,[4*] Victor V. Moshchalkov[3]

1 Research Institute of Superconductor Electronics, Nanjing University, Nanjing 210093, China

2 National Institute for Materials Science, Tsukuba 305-0044, Japan

3 INPAC-Institute for Nanoscale Physics and Chemistry, KU Leuven, Leuven B-3001, Belgium

4 Theory of Nanomaterials Group, KU Leuven, Leuven B-3001, Belgium

5 National Laboratory for Superconductivity, Institute of Physics, and Beijing National Laboratory for Condensed Matter Physics, Chinese Academy of Sciences, Beijing 100080, China

6 Wuhan National High Magnetic Field Center and School of Physics, Huazhong University of Science and Technology, Wuhan 430074, China

7 Electron Microscopy for Materials Research (EMAT), University of Antwerp, Antwerp B-2020, Belgium.

8 Synergetic Innovation Center in Quantum Information and Quantum Physics, University of Science and Technology of China, Hefei, Anhui 230026, China

9 Graduate School of Chemical Sciences and Engineering, Hokkaido University, Sapporo 060-0810, Japan

**Nematic order often breaks the tetragonal symmetry of iron-based superconductors. It arises from regular structural transition or electronic instability in the normal phase. Here, we report the observation of a nematic superconducting state, by measuring the angular dependence of the in-plane and out-of-plane magnetoresistivity of $Ba_{0.5}K_{0.5}Fe_2As_2$ single crystals. We find large twofold oscillations in the vicinity of the superconducting transition, when the direction of applied magnetic field is rotated within the basal plane. To avoid the influences from sample geometry or current flow direction, the sample was designed as Corbino-shape for in-plane and mesa-shape for out-of-plane measurements. Theoretical analysis shows that the nematic superconductivity arises from the weak mixture of the quasi-degenerate *s*-wave and *d*-wave components of the superconducting condensate, most probably induced by a weak anisotropy of stresses inherent to single crystals.**

The discovery of iron-based superconductors with high critical temperature has revived the interest for unconventional superconductivity [1]. Their similarity to high-$T_c$ cuprates [2] suggests that magnetic fluctuations might be the leading mechanism of superconducting pairing [3,4]. The peculiar feature of these materials is the essentially multiband character of pairing [4-6]. The Fermi surface of iron-based superconductors is very rich and can modified by introducing doping atoms [5,7,8] which mostly change the number of electrons in the bands near the Fermi level. Moreoer, the symmetry of the electron pairing appears to be intertwined with the topology of the Fermi surface which varies drastically across the compounds [5,7,8]. Furthermore, a crossover between different pairing types, namely, between $s_{\pm}$-wave and *d*-wave symmetry, under variation of electron/hole doping is highly debated in the community [9-11], however, consensus is not reached with several opposing parties [12-14]. At the same time, other authors claim the near-degeneracy of *s*- and *d*-wave components [15,16] or strong subdominat *d*-wave component [17] in iron-based superconductors, particularly, in $Ba_{1-x}K_xFe_2As_2$ [18].

Iron-based superconductors display an antiferromagnetic dome in the temperature-doping phase diagram, centered in the parent compound [19]. Numerous studies revealed the existence of nematic order just above the magnetic ordering phase, characterized by the tetragonal rotational symmetry breaking and always accompanied by the structural transition to orthorhombic phase [19-22]. The superconducting state nucleating at the bottom of this dome in electron- or hole-doped compounds displays nematic symmetry as well, which is testified, *e.g.*, by the $C_2$ symmetry of the superconducting gap along the Fe-Fe bond direction [23]. However, it should be mentioned that the magnetic order (no pressure) is absent in the FeSe system, and the nematic state also exists in the normal state, further researches are necessary to understand the mechanims of the nematic state in superconductivity. Electronic nematic orders are ubiquitous also in other unconventional superconductors, such as high-$T_c$ cuprates and heavy-fermion compounds [24]. They always nucleate in the normal phase, most probably via spin-fluctuation mechanism [19], and depend on the degree of doping. Thus in hole doped iron-based superconductors such as $Ba_{1-x}K_xFe_2As_2$ the nematic order is observed till $x \approx 0.3$, when the antiferromagnetic phase boundary is reached, and disappears at higher values of doping including the optimal one [25] ($x = 0.4$).

Here we present evidence of a nematic superconducting state which is not accompanied by tetragonal symmetry breaking of the lattice or onset of magnetic order. Measuring angular dependences of the in-plane magnetoresistivity in single-crystalline samples of iron-based superconductors near optimal hole doping, we find twofold oscillations with the rotation of the applied magnetic field. The $C_2$ symmetry of the superconducting condensate persists from the nucleation till low temperatures and for a range of dopings near $x = 0.5$. The reason for this unusual nematic state is found to be a mixture of quasidegenerate $s_{\pm}$-wave and *d* -wave-components of the superconducting condensate by weak anisotropy of stresses and defects distribution in single-crystalline films.

## RESULTS

**Measurement setup.** The optimally doped single crystals of $Ba_{0.5}K_{0.5}Fe_2As_2$ with a $T_c = 38$ K were grown by a high-pressure method [26]. An annular-electrode method was adopted to measure the in-plane electrical resistivity ($\rho_{ab}$) of selected single crystal (Supplementary Fig. 2); the outermost annular golden pattern was fabricated with a diameter of 80 $\mu$m (see Fig. 1). The electric current flows radially from the center to the outermost electrode as in the Corbino disc [27], thus eliminating the anisotropic Lorentz force effect. The $\rho_{ab}$ was measured with the Physical Properties Measurement System (PPMS; Quantum Design). The angle $\theta$ is defined as that between the magnetic field and the *a(b)*-axis of the lattice, as indicated in Fig. 1a. The dependence of $\rho_{ab}$ on $\theta$ was measured by rotating the *ab*-plane around the *c*-axis in a fixed magnetic field parallel to the *ab*-plane and at a fixed temperature. The potential misalignment of the magnetic field against the *ab*-plane was estimated to be less than ±1° (Supplementary Fig. 11). Variation of the sample temperature due to sample rotation was verified to be less than 0.008 K (Supplementary Fig. 12).

**In-plane magnetoresistivity**. Figs. 2a and 2b show the measured angular dependence of the in-plane magnetoresistivity $\rho_{ab}$ (IMR) in vicinity of superconducting transition under magnetic field of 9 T. The $\rho_{ab}$ displays oscillations as a function of $\theta$ below the $T_c$-onset. These oscillations disappear as the temperature approaches $T_c$-onset from below, reaching an isotropic resistivity. Away from $T_c$-onset and closer to the $T_c$-offset, a twofold oscillation of IMR is observed as a nearly sinusoidal variation, in which the maxima of IMR appears at $\theta \approx 135°$ and $315°$, and the minima at $\theta \approx 45°$ and $225°$. Fig. 3 shows the change in symmetry under various magnetic fields and temperatures more clearly. The color contours represent the normalized magnetoresistivity ($\rho-\rho_0)/\rho_n$, where $\rho$ is $\rho_{ab}$, $\rho_0$ is the $\rho_{ab}$ at $\theta = 0$, and $\rho_n$ is the normal state $\rho_{ab}$ at a temperature of 39 K. Note that the anisotropic factor can be up to -10% to 15.6% for the minima and maxima magnetoresistivity, respectively. Such anisotropic factors are dramatically larger than that of the structure distortion induced anisotropic one, which is normally around 0.1% along the *a*- and *b*-axis.

We carried out $\rho_{ab}$ measurements for more than 10 different single crystals and found that the maximum (minimum) of $\rho_{ab}(\theta)$ is always in the close neighborhood to the $(\pm 1, \pm 1, 0)$ direction of the crystal, namely, the Fe-Fe ($\Gamma$M) bond direction. To confirm that the $\rho_{ab}(\theta)$ maxima (minima) are independent of the initial direction of the applied magnetic field, similar measurements were repeated at different initial angle $\phi$ of 45° and 90° (Supplementary Fig. 13). The maximum (minimum) of $\rho_{ab}(\theta)$ was always observed along the direction of the Fe-Fe bond. When the temperature is close to the $T_c$-offset, the $\rho_{ab}$ shows a mixture of twofold and fourfold oscillations (see Supplementary Figs. 9 and 10). The fourfold oscillation, however, has a considerably lower amplitude.

Our central finding here is that the IMR indicates a twofold oscillation as a function of $\theta$ when the temperature is set near the mid-point of the superconducting transition. If the temperature is much higher than the mid-point, the oscillation disappears with a constant resistivity independent of $\theta$. When the temperature is much lower than the mid-point, on the contrary, the resistivity becomes zero for the whole range of $\theta$. Increase of the magnetic field does not affect the phase of the oscillation but makes its amplitude larger.

**Out-of-plane magnetoresistivity**. Complementarily, $\rho_c(\theta)$ was measured by using the mesa technique (well-studied on the cuprates superconductors [28]). Fig. 1b gives the scheme of the mesa structure in which the current flows from the top side of the mesa to the crystal substrate. $\mu_0 H$ was applied within the *ab*-plane and the sample was rotated around the *c*-axis as in the IMR experiments. Thus the Lorentz force effect is eliminated as previously, because $\mu_0 H$ is always perpendicular to the current flow direction. The angular-dependent $\rho_c$ in the superconducting transition region is given in Supplementary Fig.6. We can see that $\rho_c(\theta)$ demonstrates oscillations similar to $\rho_{ab}(\theta)$.

**In-plane upper critical field.** Magnetoresistivity can be linked to thermally activated vortex creep and flux flow at the offset and the onset of critical temperature, respectively, therefore an anisotropy of the pinning potential might be the reason for the observed $C_2$ oscillations of IMR [29]. To check this possibility, we extracted the second critical field $H_{c2}$, which does not depend on the pinning potential, from electrical transport measurements. Fig. 4a shows that $H_{c2}(\theta)$ displays the same anisotropic behavior, thus, excluding anisotropic distribution of the pinning sites as the source of the $\rho_{ab}(\theta)$ tetragonal symmetry breaking. We also employed the pulsed high magnetic fields (57 T) to study on the $H_{c2}(\theta)$ in relatively low temperature of 35 K (4.5 K below the $T_c$), similar two-fold anisotropic behavior (see Supplementary Figs. 7, 8 and 9) was found as those of static magnetic fields.

The procedure used here allows to measure $\rho_{ab}$ with high accuracy and low noise, however, it might induce metastable states when the field is deviated from the initial direction. To rule out this possibility, field-cooling experiments were made in which the fixed magnetic field was applied at $T = 140$ K while the temperature was lowered till the measuring value, the procedure being repeated for each direction of the applied field. The obtained angular dependencies of IMR are practically identical to the ones obtained with the previous procedure (see Supplementary Note 8), indicating the absence of metastability in our measurements.

**Technical error exclusion.** Analyzing possible reasons for tetragonal symmetry breaking of $\rho_{ab}(\theta)$, $\rho_c(\theta)$, and $H_{c2}(\theta)$, we note that according to the diagram established for bulk $Ba_{1-x}K_xFe_2As_2$ (ref. 30), the present materials ($x \approx 0.5$) are expected to be free of orthorhombic distortions or antiferromagnetic ordering. Tunneling electronic microscopy (TEM) measurements also show the absence of twinning till low temperatures (Supplementary Fig. 1). This implies that symmetry breaking mechanisms in the normal phase, like orbital and spin instabilities, can be ruled out in the present case. To rule out effect of shape, samples of different geometries, including those of circular shapes, were additionally investigated and have shown no qualitative differences for the angular dependence of IMR. Finally, we investigated the effect of misalignment of the $c$-axis of the crystal and the rotation axis on the amplitude of the IMR oscillation, and found that it is several orders of magnitude smaller than the observed ones (see the Supplemtary Fig. 11).

To summarize, the geometry of the samples and of the setup, as well as the anisotropy of vortex pinning potential and the metastability can be ruled out as possible causes for the $C_2$ anisotropy of $\rho_{ab}(\theta)$ and $\rho_c(\theta)$. The lattice remains tetragonal for the optimally doped crystal from room to low temperatures and no magnetic ordering is observed [30], which excludes usual mechanisms such as orthorhombic structural phase transition, orbital and spin instabilities as reasons for prominent two-fold symmetry of IMR in Fig. 2. Additionally, in all measurements the anisotropy of IMR can only be observed in the transition domain between the normal and superconducting state. This experimental analysis demonstrates that the observed tetragonal symmetry breaking is truly unusual, and is intrinsically intertwined with the superconducting state. Contrary to conventional nematic states reported for tetragonal superconductors [8,19-24], we face here a new manifestation of nematicity, the nematic superconducting state.

## DISCUSSION

The observed nematicity in the superconducting state could in principle arise from spontaneous symmetry breaking in the coupling between superconducting and nematic order parameter [9]. This would naturally explain why the nematic order is absent in the normal phase and only shows up in superconducting state. However, on symmetry reasons the nematic order parameter $\varphi$ can only couple simultaneously to two superconducting order parameters of different symmetry, $\Delta_s$ and $\Delta_d$, in a fashion $\propto \varphi \Delta_s \Delta_d$ [9], which leads to its equilibrium value of the form $\varphi^{(0)} \propto \Delta_s \Delta_d$. The latter is quadratic in superconducting order parameter and, therefore, negligible at the nucleation point. This contradicts $H_{c2}(\theta)$ data (Fig. 4a) which are consistent with the existence of nematic superconductivity already at the nucleation. Thus, a simple Ginzburg-Landau argument allows to rule out the spontaneous symmetry breaking as possible reason for the observed large $C_2$ oscillations of IMR and $H_{c2}(\theta)$.

We are forced, therefore, to assume that the observed nematicity of the superconducting state is triggered by tetragonal symmetry breaking present already in the normal phase, despite the lack of its evidence from structural data. To prove the existence of the tiny nematicity in the normal state, we digitalized IMR data at $T > T_c$ and found indeed their low-symmetry angular oscillation with an amplitude of 0.8% (Supplementary Fig.16 and Supplementary Table 1). This preexisting weak nematicity can induce significant tetragonal symmetry breaking in the superconducting state only in the presence of quasidegenerate components of superconducting order parameters of different symmetry, which seems indeed to be the case for $Ba_{0.5}K_{0.5}Fe_2As_2$. The current understanding is that in the hole optimally doped and underdoped range ($x \leq 0.4$), $Ba_{1-x}K_xFe_2As_2$ is likely to exhibit $s_\pm$-wave pairing [11], while in the strongly hole doped range it shows evidences of $d$-wave superconductivity, particularly for the completely doped, namely, $KFe_2As_2$ [11,31,32]. At intermediate doping levels, a crossover between these two pairing symmetries is expected [10]. Based on these facts, we performed microscopic calculations of $H_{c2}(\theta)$ taking into account possible mixing of $s_\pm$-wave and $d$-wave superconducting components. In these calculations we assumed that the samples are close to the clean limit due to a very short coherence length (~3 nm) and the existence of strong angular oscillations of the IMR and the critical field.

$H_{c2}(\theta)$ was calculated using the linearized Eilenberger equations, suitable to treat the clean limit. The Fermi surface and the Fermi velocity were obtained within the five-band model [18] in the rigid band approximation (see Supplementary Note 1 - 4 for more details). The pairing potential was parametrized by a procedure similar to Ref. 33 and Ref. 4. Magnetoresistivity in the transition domain from normal to superconducting state was calculated using Tinkham's model [34].

The parametrization of the pairing potential includes only one free parameter, the critical temperature of the secondary component ($T_{c2}$). Fixing it, the critical temperature of the main component ($T_{c1}$) is found from the known critical temperature of the system ($T_c$). As $T_{c2}$ approaches $T_{c1}$, the system becomes unstable under small perturbations that mix the two components. By admixing a small term to the pairing potential, we obtained a $H_{c2}$ and resistivity angular dependencies with similar features to the experimental curves, Fig. 4a and Fig. 2, respectively. The maxima and the minima of the simulated $H_{c2}$ and resistivity curves are along the Fe-Fe bond direction and a strong anisotropy is present despite the fact that only an arbitrary small admixed term is required to obtain this result. Similar results were obtained in the cases, $T_{c1} > T_{c2}$ and $T_{c1} < T_{c2}$. Importantly, the angle $\theta$ at which the maximum of IMR and the minimum of $H_{c2}$ arises is only defined by the symmetry of the mixing components of superconducting order parameter, $s_\pm$ and $d_{x^2-y^2}$, and does not depend on the details of the weak symmetry breaking triggering the observed nematicity. It is also independent from the extent of mixing of the $s_\pm$-wave and $d$-wave superconducting components. Finally, the simulation results depend weakly on changes in the Fermi surface and electron pockets eccentricity corresponding to small variations of the doping level.

The fact that the anisotropy direction of IMR and $H_{c2}(\theta)$ varies weakly across the set of investigated samples, and that it is reproduced by a simple model for a wide range of its parameters, is the major evidence of the validity of the proposed mechanism for nematic superconducting state. The correct anisotropy angle is obtained exclusively from the admixture of the $d_{x^2-y^2}$-wave component. An additional support for this scenario comes from a very small value of the mixing term in the pairing potential needed to reproduce the main features of angular dependence of IMR and $H_{c2}(\theta)$. For example, the necessary extent of mixing of the two superconducting components whose critical temperatures ($T_{c1}$ and $T_{c2}$) differ by 2 K would require an amplitude of the mixing term in pairing potential amounting to only 1% of the main (tetragonal) pairing potential. This is in line with the experimental fact that the preexisting Since there is no evidence for nematic order in bulk compounds with optimally or over doped compositions [30], we should admit that the weak tetragonal symmetry breaking observed here in the normal phase (Supplementary Fig.16) is rooted in their preparation procedure of the single-crystalline films. Indeed, the latter have been grown under a very high pressure (~3 GPa) [26], which explains the presence of anisotropic stresses in the $a$-$b$ plane inducing at their turn, a weak tetragonal symmetry breaking. The extrinsic reason for the latter explains naturally (*i*) the weakness of the symmetry breaking in the normal phase and (*ii*) its uniform character required for the observability of the two-fold oscillations of the IMR and $H_{c2}(\theta)$ in the whole sample. Indeed, if the tetragonal symmetry would be broken spontaneously, domains with different nematic orientations should have appeared. As a result, the IMR and $H_{c2}(\theta)$ anisotropy would have averaged to zero, leaving only four-fold oscillations. Finally, we can understand why the weak symmetry breaking in the normal phase is not detected as symmetry lowering of the lattice structure. For the sake of estimation one can relate the relative variations of the lattice constants $a$ and $b$ to the relative variations of the corresponding resistivity $\rho_a$ and $\rho_b$ (see, e.g. Fig. 4 in Ref. 33). Then for angular oscillations of resistivity of 0.4% (Supplementary Fig. 16b) we obtain for orthorombicity $(a-b)/(a+b)$ a value of ca 0.005%, apparently too small to be observed. On the same reason the tetragonal symmetry breaking will not be detected in the magnetic susceptibility as well.

To emphasize the difference between the observed nematic superconducting state and the superconductivity in the presence of conventional nematic order (arising from electronic/structural instability in the normal phase), we measured the angular-dependent IMR for underdoped single crystals $Ba_{0.75}K_{0.25}Fe_2As_2$ and $Ba_{0.8}K_{0.2}Fe_2As_2$, with doping values under the antiferromagnetic dome. The obtained angular-dependent IMR differs drastically from the one in Fig. 2 (see also in Suplementary Figs. 14 and 15). In contrast to the latter, it displays a twofold symmetry with minima and maxima aligned exactly along the $a$- and $b$-axis, respectively, following precisely the direction of orthorhombic structural distortion as expected.

Besides the described major features of $\rho_{ab}(\theta)$ and $H_{c2}(\theta)$, experimental results display two additional ones which are not captured by the two-component model. First, there is a deviation of the angle of maximum of IMR (minimum of $H_{c2}(\theta)$) from the Fe-Fe bond direction and, second, there is a broken reflection symmetry across the line passing through the two maxima of IMR. We first tried to exhauted the possibilities by changing fitting parameters, relaxing constraints and approximations within the two-component model to explain these features. We also tried to improve our model, adding higher harmonics in the potential decomposition. Finally, we hypothesize that the additional symmetry breaking could be explained by admixing a third component of superconducting order parameter of a different symmetry from the other two. Note this is only a small fine-tuning of the main result, i.e. symmetry breaking mechanism arising from the mixture of $s_{\pm}$- and $d_{x^2-y^2}$-wave components responsible for the direction of IMR lobes close to 135°. Concerning the multiband character of superconductivity in these compounds, theory [16,18] and experiment [15,17] give controversial evidence about the subdominant superconducting components. Simulations of $\rho_{ab}(\theta)$ and $H_{c2}(\theta)$ within the three-component model give for critical temperatures of the components values spanning a temperature domain of ca ~10 K (see Supplementary Table 2). We considered two symmetries for the third admixed component, $d_{xy}$ or $g$-wave, which gave similar results (Fig. 4b). Obviously, the admixture of the third component of the superconducting order parameter is supposed to occur on the same reason as the mixture of the first two.

Although the nematic superconducting state in the present compounds is ultimately induced by the tetragonal symmetry breaking in the normal phase, it differs qualitatively from ordinary nematicity observed in underdoped compounds in the orthorhombic structural phase. For the latter, the extrema of IMR curves follow strictly the orthorombicity axes $a$ and $b$ in both normal and superconducting phases (Supplementary Fig.14 and 15). This is a consequence of orthorhombic Fermi surface leading to anisotropic ($C_2$ symmetry) pairing potential and superconducting order parameter. On the contrary, the IMR lobes are found close to 135° in the superconducting phase of all compounds with $x \approx 0.5$ investigated here, pointing to the fact that IMR remains qualitatively unchanged for any form of extrinsic symmetry breaking in the normal phase. This behavior can be basically explained by the mixing of tetragonal $s_\pm$ and $d_{x^2-y^2}$ order parameters in all investigated compounds. The extrinsic symmetry breaking inducing this mixing is expected to be very weak, as testified by an almost isotropic IMR in the normal phase (Fig. 16), thus playing the role of a trigger rather than of a driving force for nematicity like in underdoped compounds.

## Methods

**Synthesis of crystals.** Single crystals of $Ba_{1-x}K_xFe_2As_2$ ($x$=0.2, 0.25, and 0.5) were grown using a high-pressure method (~3 GPa) [26]. X-ray diffraction confirmed that they have the tetragonal lattice, consistent with previous reports. The compound takes the tetragonal lattice form over the entire temperature range above 2 K [30]. For the optimal-doped crystal with nominal content $x$=0.5, although electron probe microanalysis revealed various potassium contents from 0.45 to 0.55 per formula unit, the crystals are still far from the orthorhombic and antiferromagnetic region, according to the well-established phase diagram [30]. Low temperature transmission electronic microscopy (TEM) measurements show the absence of intrinsic twin boundaries at temperature above 96 K (see Supplementary Fig. 1).

**Fabrication of Corbino samples.** A selected single crystal was cleaved over the *ab*-plane with a thickness of 1 $\mu$m. An annular-electrode method was adopted to measure the in-plane electrical resistivity ($\rho_{ab}$) (Fig. 1A); the outermost annular golden pattern was fabricated with a diameter of 80 $\mu$m (see Supplementary Fig. 2). To avoid the interface oxidation or degeneration between the crystal and conducting films, the samples were deposited on Au in an in-situ fabrication system for micro- and nano-device (AdNaNo-Tek Ltd.), where the samples were in inert (Ar) or high-vacuum condition (~$10^{-10}$ Torr). The fabrication processes of Corbino disk are given as following: a) The crystal was firstly cleaved into pieces with micrometers in thickness, by using scotch tape, and then glued on MgO substrate with the *ab*-plane parallel to the substrate surface using a thin layer of epoxy. After heat-treated the epoxy at 150 °C for 4 h, the crystals were further cleaved into 1~2 μm in thickness. To improve the electrical contact, a 120-nm Au layer was evaporated onto the sliced crystal (see Supplementary Fig. 2a). The samples were then heat-treated in vacuum at 300 °C for 24 h. b) The crystal was fabricated into a disk shape using photolithography technique, after which Ar ion milling was used to remove the crystals around the disk (see Supplementary Fig. 2b), and then the photoresist was removed using acetone. c) Using another photolithography technique to pattern circuits on the disk, and Ar-ion-milling the Au layer into circle mesa in the middle of the disk (Supplementary Fig. 2c). d) A 100-nm insulating SiO layer was then coated onto the crystal, and then remove the photoresist. Thus, the crystal was covered by the SiO layer except the Au circle mesas, here the SiO layer exhibits as dark blue as shown in Supplementary Fig. 2d. To avoid the possible shortage around the disk edge, we covered a thin layer of epoxy at edge region around the disk. e) Half part of the Au circle mesas was covered by another layer of SiO using a lift-off technique as shown in red part in Supplementary Fig. 2e. The edge of the disk was coated with a thin layer of epoxy to avoid possible charge shortage. f) The whole sample was deposited another layer of 120-nm Au layer, and the fabricated the electrodes for the annular mesa by using a further step of photolithography and argon-ion etching (see Supplementary Fig. 2f).

**Fabrication of mesa samples.** The out-of-plane magnetoresistivity ($\rho_c$) was measured by using the mesa technique on the crystal of $Ba_{0.5}K_{0.5}Fe_2As_2$ as shown in Fig. 1b. The cleaved single crystals were mounted on a MgO substrate and a 120-nm Au layer was evaporated on the crystal in the in-situ fabrication system. A series of mesas - approximately 1.5 $\mu$m in thickness with various surfaces of 10×10, 20×20, and 40×40 $\mu m^2$ - was fabricated using a photolithography technique (see Supplementary Fig. 4). A 120-nm insulating SiO layer was coated to surround the mesa edges. The electrodes were contacted on the top of the mesas by additional photolithography processes. Supplementary Fig. 5 shows the temperature dependence of $\rho_c$.

**Static fields measurements for angle-dependent magnetoresistivity.** The static field measurements are carried out in the Physical Properties Measurement System (PPMS; Quantum Design). For the Cobino samples, the electric current flows radially from the center to the outermost electrode as in the Corbino disk as shown in Fig. 1, thus eliminating the Lorentz force effect. The $\rho_{ab}$ was calculated by the equation $\rho_{ab} = 2\pi hR/\ln(r_2/r_1)$, where $h$ is the thickness of the crystal, $R$ is the resistance, and $r_1$ and $r_2$ are the radii of the two voltage terminals. Here, the static magnetic fields from 1 to 9 T are applied in PPMS.

**Pulsed high magnetic field experiments for angle-dependent magnetoresistivity.** The pulsed high magnetic fields experiments are studied in Wuhan National High Magnetic Field Center and School of Physics. Here, we selected the sample as ultrathin micro-bridges $Ba_{0.5}K_{0.5}Fe_2As_2$. The fabrication process of the micro-bridge was introduced in Ref. [35]. The corresponding pulsed high magnetic fields experiment was carried out as shown in Supplementary Fig. 7. The sample was rotated along the *c*-axis, and the magnetic field up to 57 T was applied within the *ab*-plane.

**Data availability.** All data generated or analyzed during this study are included in this published article (and its Supplementary Information Files) or are available from the corresponding author upon request. The computer code generated during the current study is available from the corresponding author on request.

## ACKNOWLEDGMENTS

We thank Drs. R. Greene, K. Kuroki, H.-H. Wen, Z. X. Shi, K. Jin, M. Miyakawa, K. Hirata, and H. Ooi for valuable discussions and suggestions. This research was supported in part by the National Natural Science Foundation of China (No. 11234006, 11227904, 61501220, 61771234, 61727805, 61521001); Jiangsu Provincial Natural Science Fund (BK20150561); Opening Project of Wuhan National High Magnetic Field Center (2015KF19); World Premier International Research Center from MEXT, Japan; a Grant-in-Aid for Scientific Research (25289233 and 25289108) from JSPS; the Funding Program for World-Leading Innovative R&D on Science and Technology (FIRST Program) from JSPS; the Advanced Low Carbon Technology Research and Development Program (ALCA) from JST; and the Methusalem Funding by the Flemish Government; the European Research Council, ERC Grant No 246791-COUNTATOMS.

## AUTHOR CONTRIBUTIONS

The authors J.L., P.J.P. and J.Y. contributed equally to this work. J.L. and J.Y. designed the experiments. J.L., H-L.F., K.Y., and E.T.-M. grew the single crystals. J.L., J.Y., Y.-Y.L., M.-P.J., D.L., M.-Y.L., T.H., H.-B.W., P.-H.W., K.Y., E.T.-M., J.V., and V.V.M. fabricated the devices and measured transport properties. J.L., Y.-Y.L., Z.-Q.L., Y.-J.L., J.-F.W., and L.L. studied on the pulsed high field measurements. X.K. and G.V.T. measured the low temperature TEM. All authors discussed the data. J.L., P.J.P., and L.F.C. proposed the model and simulated the results. J.L., P.J.P., K.Y., E.T.-M. and L.F.C. analyzed the data and prepared the manuscript.

Correspondence and requests for materials should be addressed to H.-B.W. (email: hbwang1000@gmail.com), Y.K. (Yamaura.Kazunari@nims.go.jp), and L.C. (Liviu.Chibotaru@chem.kuleuven.be).

**Competing interests:** The authors declare no competing financial interests.

**Figure legends**

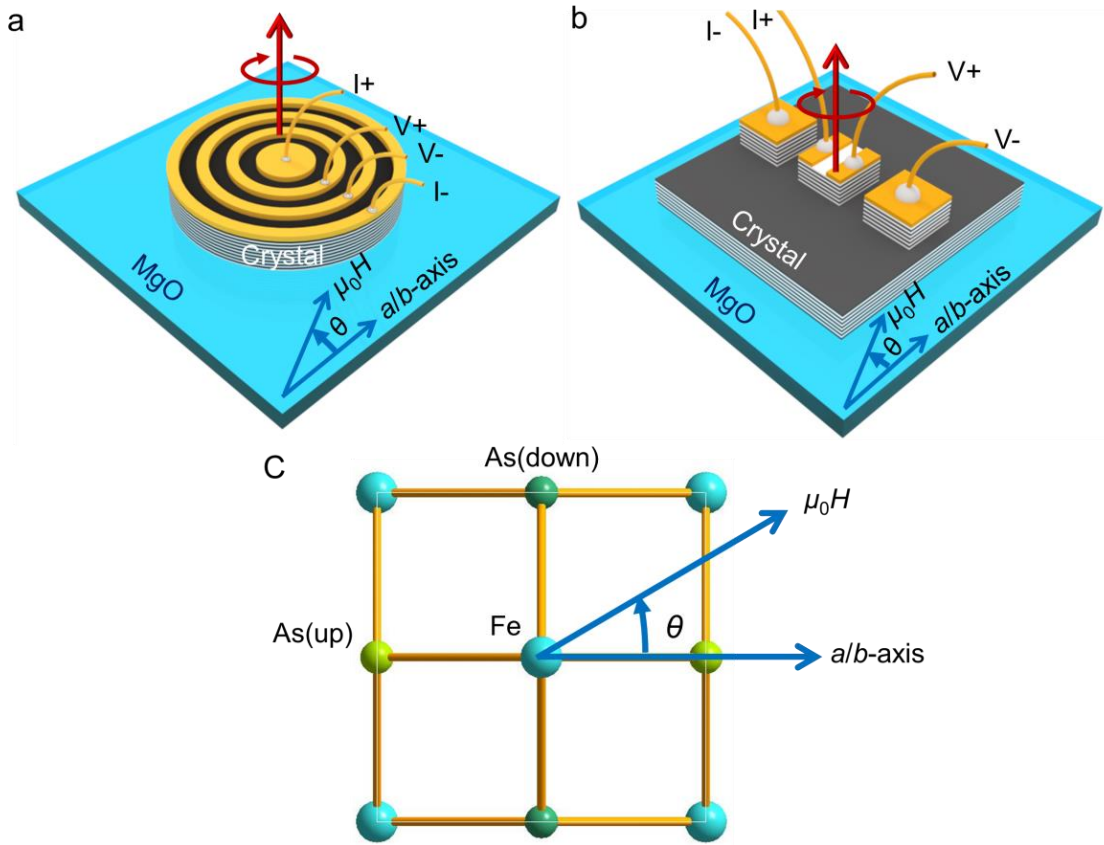

**Fig. 1 Schematic image of the sample geometry.** (a) Diagram of the Corbino-shape device for angular dependent IMR $\rho_{ab}$ measurements. The electric current is lead to flow radially from the center to the outermost electrode, where the outermost electrode was in diameter of 80 μm and the detailed sample geometry is given in Supplementary Fig.2. The magnetic field was applied to the *ab*-plane with an angular error of less than ±1°. The angle was set to zero ($\theta = 0$) when the field was parallel to one of $a(b)$-axis. Then the sample was rotated within the *ab*-plane to tune $\theta$ between $H$ and the $a(b)$-axis. (b) Diagram of the mesa device for angular dependent IMR $\rho_c$ measurements, where the thickness of the mesa was 1.5 μm (see sample geometry in Supplementary Fig.4). The magnetic field was applied within the *ab*-plane. (c) Schematic image of the rotating crystal to adjust the angle $\theta$ between $H$ and the $a(b)$-axis as well.

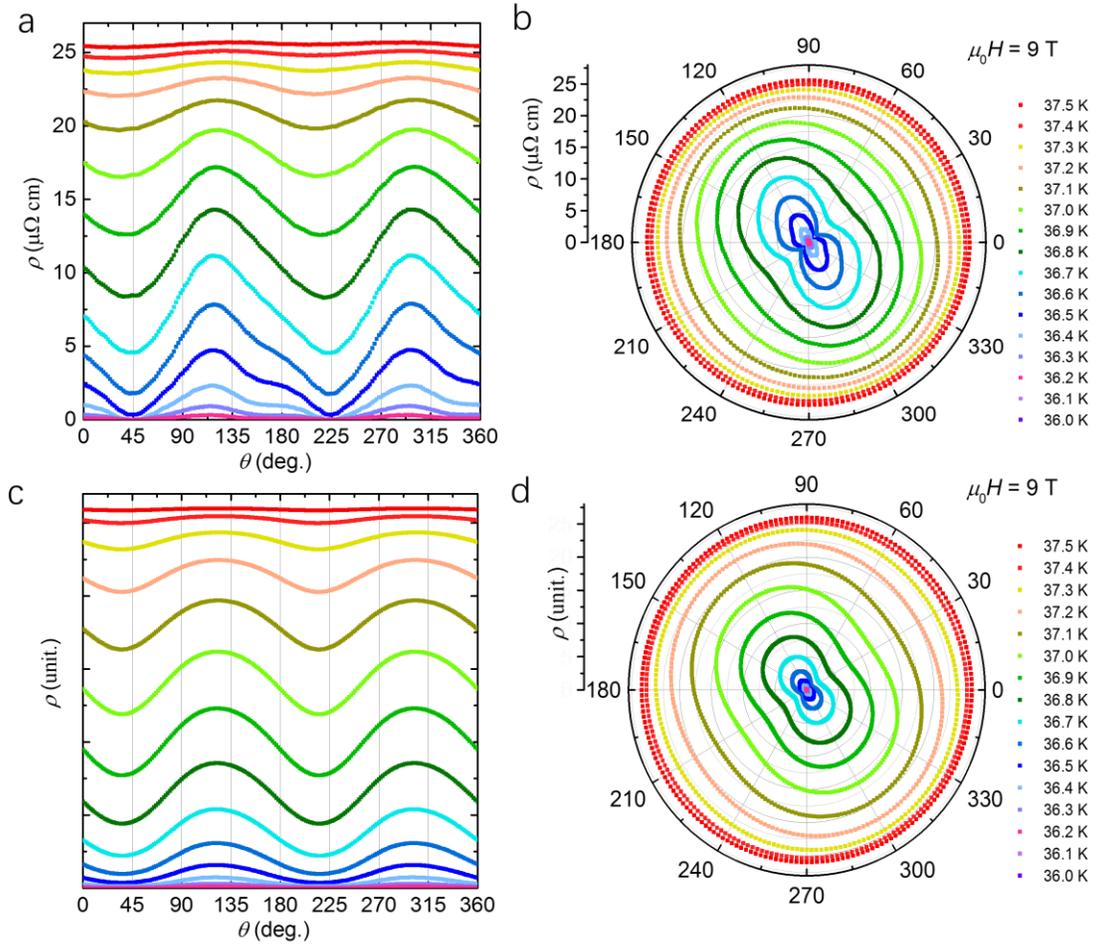

**Fig. 2 Experimental and theoretical angular dependence IMR**. Experimental (a) and theoretical (c) values of the angular dependence of the IMR, and respective polar plots of IMR experimental (b) and theoretical (d) values, at various temperatures for the applied magnetic field of 9 T for which the experimental values were obtained using the Corbino disk measurement configuration. Theoretical values correspond to the model with three components with the symmetries $s_{\pm}$, $d_{x^2-y^2}$ and $d_{xy}$.

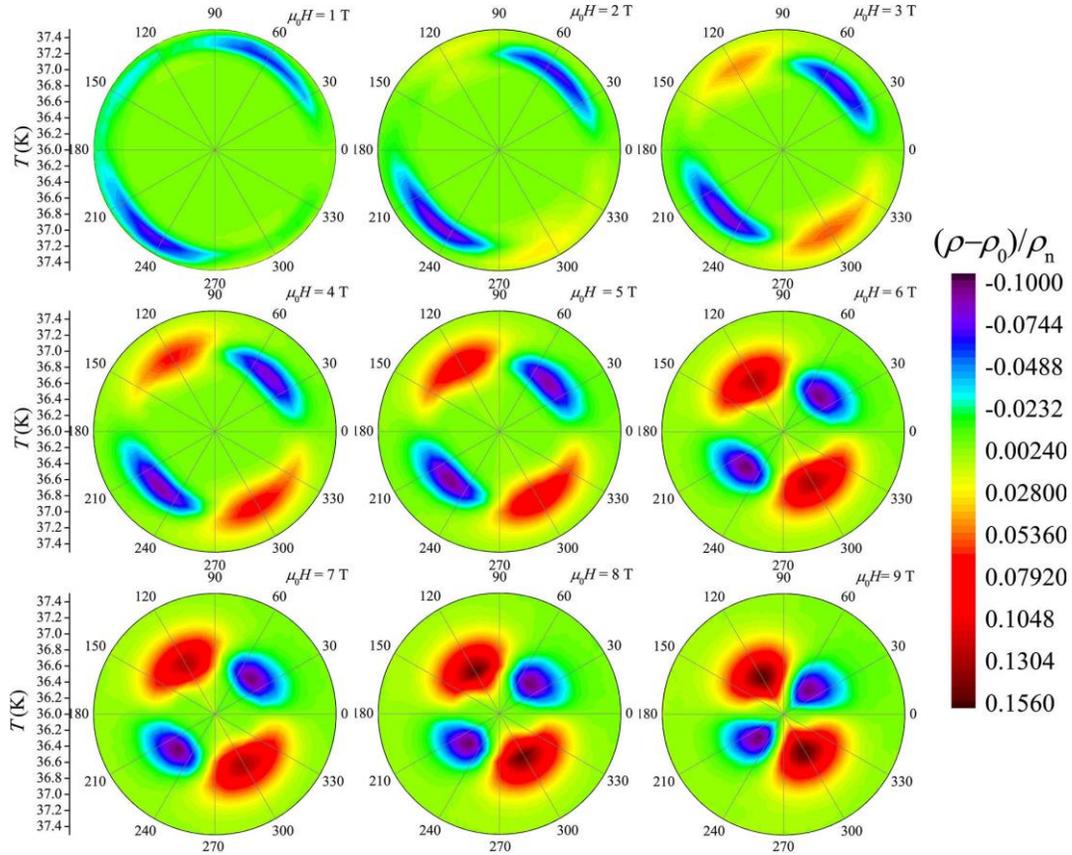

**Fig. 3 Color contours of normalized magnetoresistivity.** Angle-dependent normalized magnetoresistivity $(\rho-\rho_0)/\rho_n$ at various temperatures and magnetic fields. The color bar represents the normalized magnetoresistivity $(\rho-\rho_0)/\rho_n$, where $\rho$ is $\rho_{ab}$, $\rho_0$ is the $\rho_{ab}$ at $\theta = 0$, and $\rho_n$ is the normal state $\rho_{ab}$ at a temperature of 39 K.

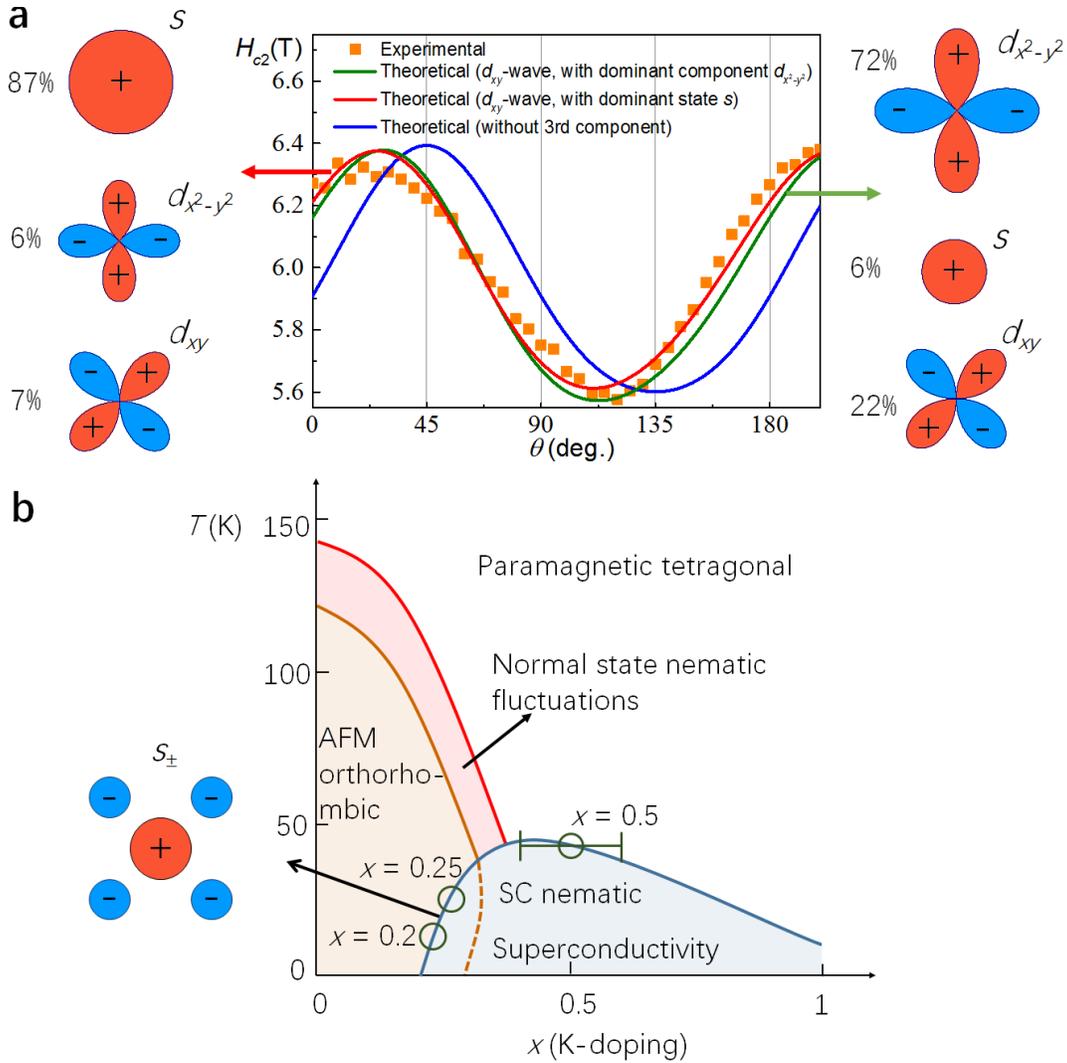

**Fig. 4 Upper critical fields and phase diagram.** (a) Angular dependence of the second magnetic critical field at 38.4 K ($T_c \approx 39$ K), retrieved from transport experiments (black filled square symbol) in the Corbino disc measurement configuration, from theoretical model with $s$-wave and $d_{x^2-y^2}$-wave symmetries (full blue line), from theoretical model with $s_\pm$-wave, $d_{x^2-y^2}$-wave and $d_{xy}$-wave symmetries (full red and blue lines depending on dominant component). The mixing of the different symmetry components of the order parameters is indicated on the left and right side panels next to a schematic representation of each component of the order parameter as function of the internal momentum of the Cooper pairs. The indicated percentages correspond to the relative weights ($r_1$, $r_2$ and $r_3$) of the wave function coefficients, $\Phi(\mathbf{k}) = r_1 \Phi_s(\mathbf{k}) + r_2 \Phi_{d_{x^2-y^2}}(\mathbf{k}) + r_3 \Phi_{d_{xy}}(\mathbf{k})$. In this schematic representation red and green indicate positive and negative value of the components, respectively. (b) Phase diagram of hole-doped $Ba_{1-x}K_xFe_2As_2$ iron-based superconductors. There is controversial evidence concerning the symmetry of the order parameter in the doping domain close to $x=1$: Refs. [12-14] argue in favor of $s_\pm$ and Ref. [11] (and references therein) in favor of $d_{x^2-y^2}$ pairing symmetry. The error bar for $x=0.5$ corresponds to 15 samples with $x$ ranging from 0.45 to 0.55. The red region demonstrates the normal state nematic fluctuations, which may originate from magnetic order, structural or charge/orbit order transition

[19]. The superconducting (SC) nematic state is observed on the basis of the present IMR results (Figs. 2 & 3).

**Supplementary Note 1. Simulation of $H_{c2}$ and magnetoresistivity using solutions of Eilenberger equations**

We used Eilenberger equations in the clean limit approximation to calculate all necessary properties of the superconducting condensate. The Eilenberger equations [3] and gap equation are

$$\left(\varepsilon_n + \frac{1}{2}\hbar \mathbf{v_F} \cdot \mathbf{\Pi}\right) f(\varepsilon_n, \mathbf{r}, \mathbf{k}) = \Delta(\mathbf{r}, \mathbf{k}) g(\varepsilon_n, \mathbf{r}, \mathbf{k}) \quad (1)$$

$$\left(\varepsilon_n + \frac{1}{2}\hbar \mathbf{v_F} \cdot \mathbf{\Pi}^*\right) f^+(\varepsilon_n, \mathbf{r}, \mathbf{k}) = \Delta^*(\mathbf{r}, \mathbf{k}) g(\varepsilon_n, \mathbf{r}, \mathbf{k}) \quad (2)$$

$$\Delta(\mathbf{r}, \mathbf{k}) = -2\pi T \sum_{n=0}^{N_c(T)} \int_{FS} V(\mathbf{k}, \mathbf{k}') f(\varepsilon_n, \mathbf{r}, \mathbf{k}') \frac{1}{(2\pi)^3 |\mathbf{v_F}|} dS_{\mathbf{k}'}, \quad (3)$$

supplemented by the equation

$$g^2(\varepsilon_n, \mathbf{r}, \mathbf{k}) + f(\varepsilon_n, \mathbf{r}, \mathbf{k}) f^+(\varepsilon_n, \mathbf{r}, \mathbf{k}) = 1, \quad (4)$$

where $N_c(T) = \Omega_{BCS}/2\pi T$ is the a cutoff in the number of states correspondent to the cutoff energy $\Omega_{BCS}$, Fermi Surface (FS) stands for Fermi surface, $dS_{\mathbf{k}'}$ is an infinitesimal element of the Fermi surface, $\varepsilon_n = (2n + 1)\pi T$ are the Matsubara frequencies, $T$ is the temperature, $\hbar$ is the reduced Planck constant, $\mathbf{\Pi} = \left(\nabla - i\frac{2\pi}{\Phi_0}\mathbf{A}\right)$, $\Phi_0$ is the magnetic flux quantum, $\mathbf{A}$ is the magnetic vector potential, $f(\varepsilon_n, \mathbf{r}, \mathbf{k})$, $f^+(\varepsilon_n, \mathbf{r}, \mathbf{k})$, and $g(\varepsilon_n, \mathbf{r}, \mathbf{k})$ are the anomalous and the conventional quasiclassical Green functions, respectively. $\Delta(\mathbf{r}, \mathbf{k})$ is the superconducting gap function that depends on the spatial position $\mathbf{r}$ and on the internal momentum $\mathbf{k}$ of the Cooper pairs, $V(\mathbf{k}, \mathbf{k}')$ is the pairing potential, and $\mathbf{v_F}$ is the Fermi velocity. $H_{c2}$ can be found by solving the linearized version of these equations (by making $f(\varepsilon_n, \mathbf{r}, \mathbf{k}) \approx f^{(1)}(\varepsilon_n, \mathbf{r}, \mathbf{k})$ and $g(\varepsilon_n, \mathbf{r}, \mathbf{k}) \approx 1$), i.e.

$$\left(\varepsilon_n + \frac{1}{2}\hbar \mathbf{v_F} \cdot \vec{\Pi}\right) f^{(1)}(\varepsilon_n, \mathbf{r}, \mathbf{k}) = \Delta(\mathbf{r}, \mathbf{k}) \quad (5)$$

$$\Delta(\mathbf{r}, \mathbf{k}) = -2\pi T \sum_{n=0}^{N_c} \int_{FS} V(\mathbf{k}, \mathbf{k}') f^{(1)}(\varepsilon_n, \mathbf{r}, \mathbf{k}') \frac{1}{(2\pi)^3 |\mathbf{v_F}|} dS_{\mathbf{k}'}. \quad (6)$$

We note that near $H_{c2}$ the induced currents are very small and can be neglected, thus, the magnetic field in the superconductor is equal to the applied magnetic field ($\mathbf{B_a}$), i.e.

$$\nabla \times \mathbf{A} = \mathbf{B_a}. \quad (7)$$

**Supplementary Note 2. Parameterization of the pairing potential**

The parameterization of the pairing potential was constructed using a procedure similar to the ones found in Ref. [4] and Ref. [5]. We decomposed the pairing potential, $V(\mathbf{k}, \mathbf{k}')$, superconducting condensate function, $\Delta(\mathbf{r}, \mathbf{k})$ and anomalous Green functions, $f(\varepsilon_n, \mathbf{r}, \mathbf{k})$ in a set of basis functions dependent on $\mathbf{k}$ that diagonalize the pairing potential, i.e.

$$V(\mathbf{k}, \mathbf{k}') = \sum_{i=1}^{N_q} \bar{g}_i \bar{\phi}_i(\mathbf{k}) \bar{\phi}_i(\mathbf{k}'), \tag{8}$$

where $N_q$ is the number of components of the superconducting condensate. We note these functions are orthogonal and normalized,

$$\langle \bar{\phi}_i(\mathbf{k}) \bar{\phi}_j(\mathbf{k}) \rangle = \delta_{ij}, \tag{9}$$

using the norm

$$\langle \phi_i(\mathbf{k}) \phi_j(\mathbf{k}) \rangle = \int_{FS} \bar{\phi}_i(\mathbf{k}) \bar{\phi}_j(\mathbf{k}) \frac{1}{(2\pi)^3 N |\mathbf{v_F}|} dS_{\mathbf{k}'}. \tag{10}$$

Near $T_c$ we can approximate the pairing potential, $V(\mathbf{k}, \mathbf{k}')$, by the products of the form $\bar{g}_j \bar{\phi}_j(\mathbf{k}) \bar{\phi}_j(\mathbf{k}')$ where $\bar{g}_j$ correspond to critical temperatures, $T_{cj}$, in the neighborhood of the material's critical temperature $T_c$. Taking into account experimental evidences found in literature, we assumed a pairing potential with two dominant components, $s_\pm$-wave and $d_{x^2-y^2}$-wave, for which we have assigned the functions, $\bar{\phi}_1(\mathbf{k})$ and $\bar{\phi}_2(\mathbf{k})$, respectively. In this case, the pairing potential can be approximated by

$$V(\mathbf{k}, \mathbf{k}') = \sum_{i=1}^{2} \bar{g}_i \bar{\phi}_i(\mathbf{k}) \bar{\phi}_i(\mathbf{k}'), \tag{11}$$

and the expansion of the superconducting condensate function by

$$\Delta(\mathbf{r}, \mathbf{k}) = \sum_{j=1}^{2} \bar{\phi}_j(\mathbf{k}) \bar{\Delta}_j(\mathbf{r}), \tag{12}$$

where $\bar{\Delta}_1(\mathbf{r})$ and $\bar{\Delta}_2(\mathbf{r})$ are the spatial components of the gap function associated with the symmetries $s_\pm$- and $d_{x^2-y^2}$-wave, respectively. Additionally, we inserted couplings between the different components that might arise from small anisotropic distribution of the doping atoms or small strains in the sample. In this case the pairing potential transforms into

$$V(\mathbf{k}, \mathbf{k}') = \sum_{i=1}^{2} \bar{g}_i \bar{\phi}_i(\mathbf{k}) \bar{\phi}_i(\mathbf{k}') + \bar{g}_{12} \bar{\phi}_1(\mathbf{k}) \bar{\phi}_2(\mathbf{k}') + \bar{g}_{12} \bar{\phi}_2(\mathbf{k}) \bar{\phi}_1(\mathbf{k}'), \tag{13}$$

where $\bar{g}_{12}$ is the scattering rate between the $s_\pm$- and $d_{x^2-y^2}$-wave components of the order parameter. $V(\mathbf{k}, \mathbf{k}')$ is no longer in the form displayed in Supplementary Eq. (8) due to the scattering between the components. However, we can transform it back into that form by taking linear combinations $\phi_1$ and $\phi_2$ of the previous $\bar{\phi}_1$ and $\bar{\phi}_2$ functions. The paring potential becomes

$$V(\mathbf{k},\mathbf{k}') = \sum_{i=1}^{2} g_i \phi_i(\mathbf{k}) \phi_i(\mathbf{k}'), \qquad (14)$$

and the superconducting condensate function maintain the same form in respect to the new variables, i.e $\Delta(\mathbf{r},\mathbf{k}) = \sum_{j=1}^{2} \phi_j(\mathbf{k}) \Delta_j(\mathbf{r})$, where the $\Delta_1(\mathbf{r})$ and $\Delta_2(\mathbf{r})$ are linear combinations of $\bar{\Delta}_1(\mathbf{r})$ and $\bar{\Delta}_2(\mathbf{r})$ and $g_1$ and $g_2$ are constants associated with new $\phi_1$ and $\phi_2$ functions, respectively. Inserting these approximations into the linearized Eilenberger equations, we obtain

$$\left(\varepsilon_n + \tfrac{1}{2}\hbar \mathbf{v_F} \cdot \mathbf{\Pi}\right) f^{(1)}(\varepsilon_n, \mathbf{r}, \mathbf{k}) = \sum_{j=1}^{2} \phi_j(\mathbf{k}) \Delta_j(\mathbf{r}) \qquad (15)$$

$$\Delta_l(\mathbf{r}) = 2\pi g_l T \sum_{n=0}^{N_c(T)} \int_{FS} \phi_l(\mathbf{k}') f^{(1)}(\varepsilon_n, \mathbf{r}, \mathbf{k}') \frac{1}{(2\pi)^3 |\mathbf{v_F}|} \mathrm{d}S_{\mathbf{k}'}. \qquad (16)$$

To obtain $\bar{\phi}_j(\mathbf{k})$ for Ba$_{1-x}$K$_x$Fe$_2$As$_2$, we expanded it in harmonic functions centered in each pocket of the Fermi surface, in a similar procedure to the one described in Ref. [5]. The selected harmonic functions are compliant with the symmetry of pocket's geometry and of the corresponding $\bar{\phi}_j(\mathbf{k})$, i.e.

$$\bar{\phi}_{1,h1}(\hat{\theta}) = \bar{\phi}_{1,h2}(\hat{\theta}) = \bar{\phi}_{1,h3}(\hat{\theta}) = 1 \qquad (17)$$

$$\bar{\phi}_{2,h1}(\hat{\theta}) = \bar{\phi}_{2,h2}(\hat{\theta}) = \bar{\phi}_{2,h3}(\hat{\theta}) = \cos(2\hat{\theta}), \qquad (18)$$

and

$$\bar{\phi}_{1,e1}(\hat{\theta}) = a_{1,1} + a_{1,2}\cos(2\hat{\theta}) \qquad (19)$$

$$\bar{\phi}_{1,e2}(\hat{\theta}) = a_{1,1} - a_{1,2}\cos(2\hat{\theta}) \qquad (20)$$

$$\bar{\phi}_{2,e1}(\hat{\theta}) = a_{2,1}\cos(2\hat{\theta}) + a_{2,2} \qquad (21)$$

$$\bar{\phi}_{2,e2}(\hat{\theta}) = a_{2,1}\cos(2\hat{\theta}) - a_{2,2}, \qquad (22)$$

where $\hat{\theta}$ is an angle defined according to a fixed lattice axis and the center of the respective pocket, $h1$, $h2$ and $h3$ correspond to the inner, the outer central hole pockets and corner hole pockets and $e1$ and $e2$ correspond to the electron pockets on the sides of the Fermi surface, all of which can be identified in Supplementary Figure 18 for the case of Ba$_{0.5}$K$_{0.5}$Fe$_2$As$_2$.

We assume that $\bar{\phi}_j(\mathbf{k})$ is evenly distributed over the Fermi surfaces of the central holes and electron pockets, which is a good approximation since the interband scattering is stronger than the intra-band scattering between the central hole pockets and the electron pockets. However, we made a broader assumption, which was to consider that $\bar{\phi}_j(\mathbf{k})$ is evenly distributed over all pockets of the Fermi surface.

For convenience is better to express $\bar{\lambda}_1 = -\bar{g}_1 N$, $\bar{\lambda}_2 = -\bar{g}_2 N$ and $\bar{\gamma}_{12} = -\bar{g}_{12} N$ as functions of $T_{c1}$ and $T_{c2}$ and $\lambda_1 = -g_1 N$ and $\lambda_2 = -g_2 N$ as functions of $T'_{c1}$ and $T'_{c2}$.

Usual BCS equation (which can be obtained from Supplementary Eq. (3)) for a homogeneous single component superconductor (this means that $\Delta = const.$ and $V(\mathbf{k}, \mathbf{k}') = g$ ) is

$$\Delta = 2\pi T \sum_{n=0}^{N_c(T_c)} \int_{FS} \frac{\lambda}{(2\pi)^3 N |\mathbf{v}_F|} \frac{\Delta}{\sqrt{\Delta^2 + \varepsilon_n^2}} dS_{\mathbf{k}'}. \tag{23}$$

By considering that $T \to T_c$ then $\Delta \to 0$ and as usual we obtain

$$\frac{1}{\lambda} = 2\pi T \sum_{n=0}^{N_c(T_c)} \frac{1}{\varepsilon_n}. \tag{24}$$

Using it, the following sum can separated into

$$2\pi T \sum_{n=0}^{N_c(T)} \frac{1}{\varepsilon_n} = 2\pi T \sum_{n=0}^{N_c(T_c)} \frac{1}{\varepsilon_n} + 2\pi T \sum_{n=N_c(T_c)}^{N_c(T)} \frac{1}{\varepsilon_n}$$

$$= \frac{1}{\lambda} + \sum_{n=N_c(T_c)}^{N_c(T)} \frac{1}{n+1/2}. \tag{25}$$

This sum can also be approximated by

$$\sum_{n=0}^{N_c(T)} \frac{1}{n+1/2} \approx \ln(N_0(T)) + 2\ln(2) + C \tag{26}$$

in the case of large $N_c(T)$, where $N_c(T) = \Omega_{BCS}/2\pi T$ and $C = 0.5772 ...$ (Euler's constant). Using the previous expression we obtain

$$\sum_{n=N_c(T_c)}^{N_c(T)} \frac{1}{n+1/2} \approx \ln\left(\frac{T_c}{T}\right). \tag{27}$$

and thus

$$2\pi T \sum_{n=0}^{N_c(T)} \frac{1}{\varepsilon_n} \approx \frac{1}{\lambda} + \ln\left(\frac{T_c}{T}\right), \tag{28}$$

from which we get

$$\frac{\Delta(\mathbf{r})}{\lambda} - 2\pi T \sum_{n=0}^{N_c(T)} \frac{\Delta(\mathbf{r})}{\varepsilon_n} = \frac{\Delta(\mathbf{r})}{\lambda} - \frac{\Delta(\mathbf{r})}{\lambda} - \ln\left(\frac{T_c}{T}\right) \Delta(\mathbf{r}) \approx -\ln\left(\frac{T_c}{T}\right) \Delta(\mathbf{r}). \tag{29}$$

Supplementary Eq. (16),

$$\Delta_l(\mathbf{r}) = 2\pi \lambda_l T \sum_{n=0}^{N_c(T)} \int_{FS} \phi_l(\mathbf{k}') f^{(1)}(\varepsilon_n, \mathbf{r}, \mathbf{k}') \frac{1}{(2\pi)^3 |\mathbf{v}_F| N} dS_{\mathbf{k}'}, \tag{30}$$

can be approximated to

$$\Delta_l(\mathbf{r}) = 2\pi \lambda_l T \left( \sum_{n=0}^{N_c(T)} \frac{\Delta_l(\mathbf{r})}{\varepsilon_n} - \sum_{n=0}^{\infty} \frac{\Delta_l(\mathbf{r})}{\varepsilon_n} + \right.$$

$$\sum_{n=0}^{\infty} \int_{FS} \phi_l(\mathbf{k}') f^{(1)}(\varepsilon_n, \mathbf{r}, \mathbf{k}') \frac{1}{(2\pi)^3 |\mathbf{v_F}| N} dS_{\mathbf{k}'} \Bigg), \tag{31}$$

and combining it with Supplementary Eq. (29) we obtain

$$\Delta_l(\mathbf{r}) \ln\left(\frac{T'_{cl}}{T}\right) = 2\pi T \left( \sum_{n=0}^{\infty} \int_{FS} \phi_l(\mathbf{k}') f^{(1)}(\varepsilon_n, \mathbf{r}, \mathbf{k}') \frac{1}{(2\pi)^3 |\mathbf{v_F}| N} dS_{\mathbf{k}'} - \sum_{n=0}^{\infty} \frac{\Delta_l(\mathbf{r})}{\varepsilon_n} \right). \tag{32}$$

For simplicity we took the same cutoff energy, $\Omega_{\text{BCS}}$, for all components.

### Supplementary Note 3. Expansion into Landau levels and $H_{c2}$ of two-component superconductors

We made a perturbation expansion of the first equation with respect to the operator $\mathbf{v_F} \cdot \mathbf{\Pi}$. The $v$-th order of perturbation of the anomalous Green function is,

$$f^{(1)}_v(\varepsilon_n, \mathbf{r}, \mathbf{k}) = \delta_{v0} \frac{\sum_{j=1}^{\infty} \phi_j(\mathbf{k}) \Delta_j(\mathbf{r})}{\varepsilon_n} - \frac{1}{2\varepsilon_n} \hbar \mathbf{v_F} \cdot \mathbf{\Pi} f^{(1)}_{v-1}(\varepsilon_n, \mathbf{r}, \mathbf{k}). \tag{33}$$

Inserting this expansion from the zero-th to the fourth order into the gap equation (Supplementary Eq. (32)), it becomes

$$\Delta_1 \ln\left(\frac{T'_{c1}}{T}\right) = -2\pi T \sum_{n=0}^{\infty} \left[ \left\langle \phi_1^2 \left(\frac{1}{2\varepsilon_n} \hbar \mathbf{v_F} \cdot \mathbf{\Pi}\right)^2 \Delta_1 \right\rangle + \left\langle \phi_1 \phi_2 \left(\frac{1}{2\varepsilon_n} \hbar \mathbf{v_F} \cdot \mathbf{\Pi}\right)^2 \Delta_2 \right\rangle + \right.$$
$$\left. + \left\langle \phi_1 \phi_2 \left(\frac{1}{2\varepsilon_n} \hbar \mathbf{v_F} \cdot \mathbf{\Pi}\right)^3 \Delta_2 \right\rangle + \left\langle \phi_1^2 \left(\frac{1}{2\varepsilon_n} \hbar \mathbf{v_F} \cdot \mathbf{\Pi}\right)^4 \Delta_1 \right\rangle + \left\langle \phi_1 \phi_2 \left(\frac{1}{2\varepsilon_n} \hbar \mathbf{v_F} \cdot \mathbf{\Pi}\right)^4 \Delta_2 \right\rangle \right], \tag{34}$$

$$\Delta_2 \ln\left(\frac{T'_{c2}}{T}\right) = -2\pi T \sum_{n=0}^{\infty} \left[ \left\langle \phi_2^2 \left(\frac{1}{2\varepsilon_n} \hbar \mathbf{v_F} \cdot \mathbf{\Pi}\right)^2 \Delta_2 \right\rangle + \left\langle \phi_1 \phi_2 \left(\frac{1}{2\varepsilon_n} \hbar \mathbf{v_F} \cdot \mathbf{\Pi}\right)^2 \Delta_1 \right\rangle + \right.$$
$$\left. + \left\langle \phi_1 \phi_2 \left(\frac{1}{2\varepsilon_n} \hbar \mathbf{v_F} \cdot \mathbf{\Pi}\right)^3 \Delta_1 \right\rangle + \left\langle \phi_2^2 \left(\frac{1}{2\varepsilon_n} \hbar \mathbf{v_F} \cdot \mathbf{\Pi}\right)^4 \Delta_2 \right\rangle + \left\langle \phi_1 \phi_2 \left(\frac{1}{2\varepsilon_n} \hbar \mathbf{v_F} \cdot \mathbf{\Pi}\right)^4 \Delta_1 \right\rangle \right], \tag{35}$$

Afterwards, we expressed the operator $\mathbf{v_F} \cdot \mathbf{\Pi}$ in terms of ladder operators associated with the Landau levels of $\Delta_1$ and $\Delta_2$, i.e.

$$\mathbf{v_F} \cdot \mathbf{\Pi} \Delta_l(\mathbf{r}) = \frac{1}{\sqrt{2} l_c} \left( \mathbf{v}^*_{F,l} a_l - \mathbf{v}_{F,l} a_l^\dagger \right) \Delta_l(\mathbf{r}), \tag{36}$$

where $l_c = \sqrt{\Phi_0 / 2\pi B}$,

$$\mathbf{v}_{F,l} = c_{2,l} v_{Fx} + i c_{1,l} v_{Fy}, \tag{37}$$

and $a_l$ and $a_l^\dagger$ lower and raise one Landau level in the $l$-th component, i.e.

$$a_l |n\rangle_l = \sqrt{n} |n-1\rangle_l,$$
$$a_l^\dagger |n\rangle_l = \sqrt{n+1} |n+1\rangle_l.$$

These ladder operators are related to $\mathbf{\Pi}$ by

$$\begin{pmatrix} a_1 \\ a_1^\dagger \end{pmatrix} = \frac{l_c}{\sqrt{2}} \begin{pmatrix} c_{1,1} & ic_{2,1} \\ -c_{1,1}^* & ic_{2,1}^* \end{pmatrix} \begin{pmatrix} \Pi_x \\ \Pi_y \end{pmatrix}, \quad (38)$$

$$\begin{pmatrix} a_2 \\ a_2^\dagger \end{pmatrix} = \frac{l_c}{\sqrt{2}} \begin{pmatrix} c_{1,2} & ic_{2,2} \\ -c_{1,2}^* & ic_{2,2}^* \end{pmatrix} \begin{pmatrix} \Pi_x \\ \Pi_y \end{pmatrix}. \quad (39)$$

The coordinate system $(x, y, z)$ is chosen such that $\mathbf{z}$ is along $\mathbf{H}$. A coordinate system, independent of $\mathbf{H}$, can be chosen along $a$, $b$ and $c$-axes, where the coordinates along these directions are $(X, Y, Z)$. In this new coordinate system, the previous Fermi velocity components can be written as

$$v_{Fx} = v_{FX}\cos(\varphi)\cos(\theta) + v_{FY}\cos(\varphi)\sin(\theta) - v_{FZ}\sin(\varphi) \quad (40)$$

$$v_{Fy} = -v_{FX}\sin(\theta) + v_{FY}\cos(\theta), \quad (41)$$

where $\theta$ and $\phi$ are the angles such that $\mathbf{H} = |\mathbf{H}|(\sin(\varphi)\cos(\theta), \sin(\varphi)\sin(\theta), \cos(\varphi))$.

The previous system of equations expanded in the Landau levels becomes

$$-w_{(0,0),1}\Delta_1 = B[-w_{(2,0),1}(\phi_1^2)A_{(1,1),1} + w_{(2,2),1}(\phi_1^2)A^*_{(1,0),1} + w^*_{(2,2),1}(\phi_1^2)A_{(1,0),1}]\Delta_1 +$$
$$B[-w_{(2,0),2}(\phi_1\phi_2)A_{(1,1),2} + w_{(2,2),2}(\phi_1\phi_2)A^*_{(1,0),2} + w^*_{(2,2),2}(\phi_1\phi_2)A_{(1,0),2}]\Delta_2 +$$
$$B^{3/2}[-w_{3,1}(\phi_1\phi_2)A_{(2,1,1),l} + w_{3,2}(\phi_1\phi_2)A_{(1,2,1),l} + w_{3,3}(\phi_1\phi_2)A_{(1,2,2),l} -$$
$$w_{3,4}(\phi_1\phi_2)A_{(0,3),l} + w_{3,5}(\phi_1\phi_2)A_{(3,0),l} - w_{3,6}(\phi_1\phi_2)A_{(2,1,2),l}]\Delta_2 +$$
$$B^2[w_{(4,0),1}(\phi_1^2)A_{(2,2),1} - w_{(4,2),1}(\phi_1^2)A^*_{(3,1),1} - w^*_{(4,2),1}(\phi_1^2)A_{(3,1),1} +$$
$$w_{(4,4),1}(\phi_1^2)A^*_{(4,0),1} + w^*_{(4,4),1}(\phi_1^2)A_{(4,0),1}]\Delta_1 + B^2[w_{(4,0),2}(\phi_1\phi_2)A_{(2,2),2} -$$
$$w_{(4,2),2}(\phi_1\phi_2)A^*_{(3,1),2} - w^*_{(4,2),2}(\phi_1\phi_2)A_{(3,1),2} + w_{(4,4),2}(\phi_1\phi_2)A^*_{(4,0),2} +$$
$$w^*_{(4,4),2}(\phi_1\phi_2)A_{(4,0),2}]\Delta_2 \quad (42)$$

$$-w_{(0,0),2}\Delta_2 = B[-w_{(2,0),2}(\phi_2^2)A_{(1,1),2} + w_{(2,2),2}(\phi_2^2)A^*_{(1,0),2} + w^*_{(2,2),2}(\phi_2^2)A_{(1,0),2}]\Delta_2 +$$
$$B[-w_{(2,0),1}(\phi_1\phi_2)A_{(1,1),1} + w_{(2,2),1}(\phi_1\phi_2)A^*_{(1,0),1} + w^*_{(2,2),1}(\phi_1\phi_2)A_{(1,0),1}]\Delta_1 +$$
$$B^{3/2}[-w_{3,1}(\phi_1\phi_2)A_{(2,1,1),l} + w_{3,2}(\phi_1\phi_2)A_{(1,2,1),l} + w_{3,3}(\phi_1\phi_2)A_{(1,2,2),l}S -$$
$$w_{3,4}(\phi_1\phi_2)A_{(0,3),l} + w_{3,5}(\phi_1\phi_2)A_{(3,0),l} - w_{3,6}(\phi_1\phi_2)A_{(2,1,2),l}]\Delta_1 +$$
$$B^2[w_{(4,0),2}(\phi_2^2)A_{(2,2),2} - w_{(4,2),2}(\phi_2^2)A^*_{(3,1),2} - w^*_{(4,2),2}(\phi_2^2)A_{(3,1),2} +$$
$$w_{(4,4),2}(\phi_2^2)A^*_{(4,0),2} + w^*_{(4,4),2}(\phi_2^2)A_{(4,0),2}]\Delta_2 +$$
$$B^2[w_{(4,0),1}(\phi_1\phi_2)A_{(2,2),1} - w_{(4,2),1}(\phi_1\phi_2)A^*_{(3,1),1} - w^*_{(4,2),1}(\phi_1\phi_2)A_{(3,1),1} +$$

$$w_{(4,4),1}(\phi_1\phi_2)A^*_{(4,0),1} + w^*_{(4,4),1}(\phi_1\phi_2)A_{(4,0),1}]\Delta_1. \tag{43}$$

where

$$A_{(2,1,1),l} = (a_l a_l^\dagger + a_l^\dagger a_l)a_l + a_l(a_l a_l^\dagger + a_l^\dagger a_l)$$

$$A_{(1,2,1),l} = (a_l a_l^\dagger + a_l^\dagger a_l)a_l^\dagger + a_l^\dagger(a_l a_l^\dagger + a_l^\dagger a_l)$$

$$A_{(1,2,2),l} = (a_l^\dagger)^2 a_l + a_l(a_l^\dagger)^2$$

$$A_{(0,3),l} = (a_l^\dagger)^2 a_l^\dagger + a_l^\dagger(a_l^\dagger)^2]$$

$$A_{(3,0),l} = a_l^2 a_l + a_l a_l^2$$

$$A_{(2,1,2),l} = a_l^2 a_l^\dagger + a_l^\dagger a_l^2$$

$$A_{(1,3),l} = a_l a_l^\dagger a_l^\dagger a_l^\dagger + a_l^\dagger a_l^\dagger a_l^\dagger a_l + a_l^\dagger a_l^\dagger a_l a_l^\dagger + a_l^\dagger a_l a_l^\dagger a_l$$

$$A_{(3,1),l} = a_l a_l^\dagger a_l a_l + a_l^\dagger a_l a_l a_l + a_l a_l a_l a_l^\dagger + a_l a_l a_l^\dagger a_l$$

$$A_{(2,2),l} = a_l a_l^\dagger a_l^\dagger a_l + a_l^\dagger a_l a_l a_l^\dagger + a_l a_l^\dagger a_l a_l^\dagger + a_l^\dagger a_l a_l^\dagger a_l$$

$$+ a_l a_l a_l^\dagger a_l^\dagger + a_l^\dagger a_l^\dagger a_l a_l$$

$$A_{(0,4),l} = a_l^\dagger a_l^\dagger a_l^\dagger a_l^\dagger$$

$$A_{(4,0),l} = a_l a_l a_l a_l$$

$$A_{(1,1),l} = a_l a_l^\dagger + a_l^\dagger a_l$$

$$A_{(1,0),l} = a_l^2$$

$$A_{(0,1),l} = (a_l^\dagger)^2,$$

and

$$w_{(2,2),l}(f(\mathbf{k})) = \sum_{n=0}^{\infty} \frac{(2\pi)^2 T\hbar^2}{2^3 \varepsilon_n^3 \Phi_0} \langle f(\mathbf{k})(\bar{v}_{F,l})^2 \rangle$$

$$w^*_{(2,2),l}(f(\mathbf{k})) = \sum_{n=0}^{\infty} \frac{(2\pi)^2 T\hbar^2}{2^3 \varepsilon_n^3 \Phi_0} \langle f(\mathbf{k})(\bar{v}^*_{F,l})^2 \rangle$$

$$w_{(2,0),l}(f(\mathbf{k})) = \sum_{n=0}^{\infty} \frac{(2\pi)^2 T\hbar^2}{2^3 \varepsilon_n^3 \Phi_0} \langle f(\mathbf{k})|\bar{v}_{F,l}|^2 \rangle$$

$$w_{(3,1),l}(f(\mathbf{k})) = \sum_{n=0}^{\infty} \frac{(2\pi)^{5/2} T\hbar^3}{2^{9/2} \varepsilon_n^4 \Phi_0^{3/2}} \langle f(\mathbf{k})|\bar{v}_{F,l}|^2 \bar{v}^*_{F,l} \rangle$$

$$w_{(3,2),l}(f(\mathbf{k})) = \sum_{n=0}^{\infty} \frac{(2\pi)^{5/2}T\hbar^3}{2^{9/2}\varepsilon_n^4 \Phi_0^{3/2}} \langle f(\mathbf{k})|\bar{v}_{F,l}|^2 \bar{v}_{F,l}\rangle$$

$$w_{(3,3),l}(f(\mathbf{k})) = \sum_{n=0}^{\infty} \frac{(2\pi)^{5/2}T\hbar^3}{2^{9/2}\varepsilon_n^4 \Phi_0^{3/2}} \langle f(\mathbf{k})(\bar{v}_{F,l})^2 \bar{v}_{F,l}^*\rangle$$

$$w_{(3,4),l}(f(\mathbf{k})) = \sum_{n=0}^{\infty} \frac{(2\pi)^{5/2}T\hbar^3}{2^{9/2}\varepsilon_n^4 \Phi_0^{3/2}} \langle f(\mathbf{k})(\bar{v}_{F,l})^2 \bar{v}_{F,l}\rangle$$

$$w_{(3,5),l}(f(\mathbf{k})) = \sum_{n=0}^{\infty} \frac{(2\pi)^{5/2}T\hbar^3}{2^{9/2}\varepsilon_n^4 \Phi_0^{3/2}} \langle f(\mathbf{k})(\bar{v}_{F,l}^*)^2 \bar{v}_{F,l}^*\rangle$$

$$w_{(3,6),l}(f(\mathbf{k})) = \sum_{n=0}^{\infty} \frac{(2\pi)^{5/2}T\hbar^3}{2^{9/2}\varepsilon_n^4 \Phi_0^{3/2}} \langle f(\mathbf{k})(\bar{v}_{F,l}^*)^2 \bar{v}_{F,l}\rangle$$

$$w_{(4,0),l}(f(\mathbf{k})) = \sum_{n=0}^{\infty} \frac{(2\pi)^3 T\hbar^4}{2^6 \varepsilon_n^5 \Phi_0^2} \langle f(\mathbf{k})|\bar{v}_{F,l}|^4\rangle$$

$$w_{(4,2),l}(f(\mathbf{k})) = \sum_{n=0}^{\infty} \frac{(2\pi)^3 T\hbar^4}{2^6 \varepsilon_n^5 \Phi_0^2} \langle f(\mathbf{k})|\bar{v}_{F,l}|^2 (\bar{v}_{F,l})^2\rangle$$

$$w_{(4,2),l}^*(f(\mathbf{k})) = \sum_{n=0}^{\infty} \frac{(2\pi)^3 T\hbar^4}{2^6 \varepsilon_n^5 \Phi_0^2} \langle f(\mathbf{k})|\bar{v}_{F+}|^2 (\bar{v}_{F,l}^*)^2\rangle$$

$$w_{(4,4),l}(f(\mathbf{k})) = \sum_{n=0}^{\infty} \frac{(2\pi)^3 T\hbar^4}{2^6 \varepsilon_n^5 \Phi_0^2} \langle f(\mathbf{k})(\bar{v}_{F,l})^4\rangle$$

$$w_{(4,4),l}^*(f(\mathbf{k})) = \sum_{n=0}^{\infty} \frac{(2\pi)^3 T\hbar^4}{2^6 \varepsilon_n^5 \Phi_0^2} \langle f(\mathbf{k})(\bar{v}_{F,l}^*)^4\rangle$$

$$w_{(0,0),l} = \ln\left(\frac{T'_{cl}}{T}\right).$$

The values of variables $c_{1,1}$, $c_{2,1}$, $c_{1,2}$ and $c_{2,2}$ are arbitrary and we have chosen these such that $w_{2,2}(\phi_1^2) = 0$ and $w_{2,2}(\phi_2^2) = 0$. Projecting the first and second equation into the lowest Landau levels $\Delta_1^0$ and $\Delta_2^0$, respectively, we simplified the previous equations to

$$0 = w_{(0,0),1} - B_{c2}[w_{(2,0),1}(\phi_1^2)] - B_{c2}[w_{(2,0),2}(\phi_1\phi_2)_1 \langle 0|0\rangle_2]$$
$$+ 3B_{c2}^2[w_{(4,0),1}(\phi_1^2)] + 3B_{c2}^2[w_{(4,0),2}(\phi_1\phi_2)_1 \langle 0|0\rangle_2] \quad (44)$$

$$0 = w_{(0,0),2} - B_{c2}[w_{(2,0),2}(\phi_2^2)] - B_{c2}[w_{(2,0),1}(\phi_1\phi_2)_2 \langle 0|0\rangle_1]$$
$$+ 3B_{c2}^2[w_{(4,0),2}(\phi_2^2)] + 3B_{c2}^2[w_{(4,0),1}(\phi_1\phi_2)_2 \langle 0|0\rangle_1], \quad (45)$$

where $|0\rangle_1$ and $|0\rangle_2$ are the Dirac ket's directly corresponding to the Landau levels of $\Delta_1^0$ and $\Delta_2^0$, respectively, and $B$ was replaced by $B_{c2}$ since, by definition, it is the second critical field corresponding to a given $T$.

## Supplementary Note 4. $H_{c2}$ of three-component superconductors

The experimental curves display other features rather that the strong anisotropy close to Fe-Fe bond direction that cannot be explained within the two-component model constructed previously. To refined our model we add a third component $\bar{\phi}_3(\mathbf{k})$ with a symmetry different from the one on $\bar{\phi}_1(\mathbf{k})$ and $\bar{\phi}_2(\mathbf{k})$ into the pairing potential which becomes

$$V(\mathbf{k},\mathbf{k}') = \sum_{i=1}^{3} \bar{g}_i \bar{\phi}_i(\mathbf{k})\bar{\phi}_i(\mathbf{k}') + \sum_{i=1}^{3}\sum_{j=1}^{3} \bar{g}_{ij}\bar{\phi}_i(\mathbf{k})\bar{\phi}_j(\mathbf{k}') \tag{46}$$

and can be transformed into the form

$$V(\mathbf{k},\mathbf{k}') = \sum_{i=1}^{3} g_i \phi_i(\mathbf{k})\phi_i(\mathbf{k}'), \tag{47}$$

as in the case of two-components. The symbols $\bar{g}_{ij}$ are the scattering constants between the $i$ component and the $j$ component, also $\bar{g}_{ij} = \bar{g}_{ji}$ and $\bar{\phi}_1(\mathbf{k}), \bar{\phi}_2(\mathbf{k})$ are functions given by Supplementary Eqs. (17)-(22). In the case of third component is of $g$-wave symmetry, $\bar{\phi}_3(\mathbf{k})$ is given by

$$\bar{\phi}_{3,h1}(\hat{\theta}) = \sin(4\hat{\theta}) \tag{48}$$

$$\bar{\phi}_{3,e1}(\hat{\theta}) = a_{3,1}\sin(4\hat{\theta}) + a_{3,2}\sin(2\hat{\theta}) \tag{49}$$

$$\bar{\phi}_{3,e1}(\hat{\theta}) = a_{3,1}\sin(4\hat{\theta}) + a_{3,2}\sin(2\hat{\theta}) \tag{50}$$

and, in the case of $d_{xy}$-wave symmetry, $\bar{\phi}_3(\mathbf{k})$ is given by

$$\bar{\phi}_{3,h1}(\hat{\theta}) = \sin(2\hat{\theta}) \tag{51}$$

$$\bar{\phi}_{3,e1}(\hat{\theta}) = a_{3,1}\sin(2\hat{\theta}) + a_{3,2}\sin(4\hat{\theta}) \tag{52}$$

$$\bar{\phi}_{3,e2}(\hat{\theta}) = a_{3,1}\sin(2\hat{\theta}) - a_{3,2}\sin(4\hat{\theta}). \tag{53}$$

Proceeding a similar way to the two component case presented above we can obtain a system of equations equivalent to Supplementary Eqs. (36) and Eq. (37) for the three-component case. The new equations are

$$0 = w_{(0,0),1} - B_{c2}[w_{(2,0),1}(\phi_1^2)] - B_{c2}[w_{(2,0),2}(\phi_1\phi_2)_1\langle 0|0\rangle_2] -$$
$$B_{c2}[w_{(2,0),3}(\phi_1\phi_3)_1\langle 0|0\rangle_3] + 3B_{c2}^2[w_{(4,0),1}(\phi_1^2)] + 3B_{c2}^2[w_{(4,0),2}(\phi_1\phi_2)_1\langle 0|0\rangle_2] +$$
$$3B_{c2}^2[w_{(4,0),3}(\phi_1\phi_3)_1\langle 0|0\rangle_3] \tag{54}$$

$$0 = w_{(0,0),2} - B_{c2}[w_{(2,0),2}(\phi_2^2)] - B_{c2}[w_{(2,0),1}(\phi_1\phi_2)_2\langle 0|0\rangle_1] -$$
$$B_{c2}[w_{(2,0),3}(\phi_3\phi_2)_2\langle 0|0\rangle_3] + 3B_{c2}^2[w_{(4,0),2}(\phi_2^2)] + 3B_{c2}^2[w_{(4,0),1}(\phi_1\phi_2)_2\langle 0|0\rangle_1] +$$
$$3B_{c2}^2[w_{(4,0),3}(\phi_3\phi_2)_2\langle 0|0\rangle_3] \tag{55}$$

$$0 = w_{(0,0),3} - B_{c2}[w_{(2,0),3}(\phi_3^2)] - B_{c2}[w_{(2,0),2}(\phi_3\phi_2)_3\langle 0|0\rangle_2] -$$
$$B_{c2}[w_{(2,0),1}(\phi_3\phi_1)_3\langle 0|0\rangle_1] + 3B_{c2}^2[w_{(4,0),3}(\phi_3^2)] + 3B_{c2}^2[w_{(4,0),2}(\phi_3\phi_2)_3\langle 0|0\rangle_2 +$$
$$3B_{c2}^2[w_{(4,0),1}(\phi_3\phi_1)_3\langle 0|0\rangle_1], \tag{56}$$

where $|0\rangle_3$ is the Dirac ket's corresponding to the lowest Landau level of the new component.

**Supplementary Note 5. Resistivity model**

The resistivity of high temperature superconductors in the vicinity of the normal - superconducting phase boundary, mainly originates from thermally activated vortex creep motion in the offset of the transition between normal to superconducting state and thermally activated flux flow in the onset of the transition region. We note that the *R-T* curves of our compound resemble the ones from copper-based superconductors. To simulate these experimental results, we have firstly considered Anderson-Kim's and Tinkham's models. Anderson-Kim's model models thermally activated vortex creep motion and thus can only be applied to the offset of the transition region. Tinkham's model was in the past used to fit the full normal-superconductor transition curve in copper based superconductors very successfully. We use this model here as an interpolatory model to fit to the full curve. We note that the model for magnetoresistivity as only a phenomenological character and it's details are not of critical importance for studying the qualitative features of its angular dependence. Within Anderson-Kim's and Tinkham's models, it is required to calculate the pinning barrier energy that is proportional to the free energy density of the system. Moreover, the free energy in single-band superconductors can be related to the second critical magnetic field using the Abrikosov expression [6]. In Anderson-Kim's model, the expression for resistivity is given by

$$R = \omega_0 e^{-\frac{U_0}{k_B T}}, \tag{57}$$

where $\omega_0$ is the characteristic frequency of the flux-line vibration (from $10^5$ to $10^{11} s^{-1}$), and $U_0$ is its activation energy (or barrier's height), $k_B$ is the Boltzmann constant, and $T$ is the temperature. In Tinkham's model, the resistivity is given by

$$\frac{R}{R_n} = \left[I_0\left(\frac{U_0}{k_B T}\right)\right]^{-2} \tag{58}$$

where $R_n$ is the resistivity of the normal state, $I_0$ is the modified Bessel function, and $U_0$ is the activation energy (or barrier's height). Both these models were applied by us to capture the phenomenological features of the experimental magnetoresistivity curves. However except in this subsection, we only considered Tinkham's model within the rest of the publication (including all presented plots of magnetoresistivity) since it gives a better fitting to experimental results.

In both models, $U_0$ can be expressed in terms of the free energy and characteristic lengths, i.e., $U_0 \propto \Delta G \frac{\Phi_0}{B} \xi(T)$ where $\Delta G$ is the Gibbs free energy, $\xi$ is the coherence length of the condensate, $\Phi_0$ is the magnetic flux quantum, and $B$ is the magnitude of the induced magnetic field. We note that $\frac{\Phi_0}{B}$ is the area of a flux line (where the magnitude of the induced magnetic field can be approximated by the magnitude of the applied magnetic field, $H_a$) and $\xi$ is inserted in this expression as an average length of the jump made by the magnetic fluxes.

$\Delta G$ was calculated in phenomenological form using the Abrikosov expression,

$$\Delta G = -\frac{(B_{c2}-B)^2}{8\pi(1+(2\kappa^2-1)\beta_A)} \approx -\frac{B_{c2}^2}{8\pi(1+(2\kappa^2-1)\beta_A)}, \tag{59}$$

where $B_{c2}$ is the magnetic induction associated with the second critical magnetic field $H_{c2}$ (note that we are using the CGS unit system where $H_{c2} = B_{c2}$ if we neglect the magnetic field developed by the induced currents), $B$ is the magnitude of the magnetic field, $\kappa$ is a characteristic parameter of the material, and $\beta_A$ is the Abrikosov parameter that for a triangular lattice is $\beta_A = 1.16$. This expression is valid near $H_{c2}$, and it is obtained by considering first perturbation in the wave function and in the magnetic field to the solution of linearized Ginzburg-Landau (GL) equation in which the magnetic field is equal to the applied magnetic field as-described in the sections above. We note that we do not observe in our compound a shift in onset value of the transition (upper part of the resistivity curves in the normal-superconducting state transition) with changes in the magnetic field magnitude, as it is expected in conventional superconductors. The same behavior is observed in other high $T_c$ superconductors like copper-based superconductors. To take this aspect into account, we made one last approximation in Abrikosov expression, presented in Supplementary Eq. (59), as it is done in Ref. [7].

## Supplementary Note 6. Simulation results of $H_{c2}$ and magnetoresistivity and additional details

Supplementary Figure 18 shows $H_{c2}$ for simulated and experimental values along different directions (identified by the angle $\theta$) of the applied magnetic field. The simulated curves were calculated for two components $s_\pm$-wave and $d_{x^2-y^2}$-wave symmetry alone or with a third component with $g$-wave or $d_{xy}$-wave symmetry. Both choices for third component can explain the shifting of the maximum. Supplementary Figure 20 shows the magnetoresistivity curves of the experimental curves and the simulated curves with the same components as the $H_{c2}$ plots. Both simulations, with $g$-wave and $d_{xy}$-wave symmetries, display a magnetoresistivity with similar characteristics to the experimental curve, i.e. the maxima of the simulated curves are close to the experimental one and the mirror symmetry along the maxima is broken as it is in experiment. The simulations were conducted as described in the main text, with additional details pointed out below.

Fermi surface and Fermi velocities where obtained using tight binding five band model [8] in the rigid band approximation using Fermi level $\mu = 0.187$ eV corresponding to $n_e = 5.5$ and $x = 0.5$ in Ba$_{1-x}$K$_x$Fe$_2$As$_2$, and consistent with Fermi levels of Ref. [9].

To calculate $H_{c2}$, we solved the previous system of equations numerically, considering only the solution in which $B_{c2} \to 0$ when $T \to T'_{cm}$, where $T'_{cm}$ is the highest critical temperature of $T'_{c1}$, $T'_{c2}$, $T'_{c3}$. Using this solution, we have then expanded it into a Taylor series in the applied temperature, i.e. $B_{c2} = B_1(\theta)(1 - T/T_c) + B_2(\theta)(1 - [T/T_c]^2) + \cdots$, and considered only the first two terms of the expansion. In $H_{c2}$ plots of the Ref. [10], we can see that the terms of the form $(1 - [T/T_c]^2)$ are important, and we point out that they reflect the multiband/multicomponent nature of our compound. Calculating $H_{c2}$ for just one $s$-wave component using the same equations, i.e.

$$B_{c2} = \frac{\ln\left(\frac{T_c}{T}\right)}{\sum_{n=0}^{\infty} \frac{(2\pi)^2 T_c \hbar^2}{2^3 \varepsilon_n^3 \Phi_0} \langle |\mathbf{v_F}|^2 \rangle} \tag{60}$$

$$\approx \frac{\ln\left(\frac{T_c}{T}\right)}{\sum_{n=0}^{\infty} \frac{(2\pi)^2 T_c \hbar^2}{2^3 \varepsilon_n^3 \Phi_0} \langle |\widehat{\mathbf{v_F}}|^2 \rangle} \tag{61}$$

near the limit $T \to T_c$ the critical field is mainly a linear function of the temperature, i.e. $B_{c2} \propto (1 - T/T_c)$. Second critical field is highly underestimated in our compound if only of a single component is taken under consideration, forcing us to conclude that the high $H_{c2}$ near $T_c$ in this compound is an effect of the multiband/multicomponent nature of the material. Despite the fact the simulated $H_{c2}$ is much closer to the experimental value when considering three components, we still cannot obtain the correct absolute value of $H_{c2}$, probably due to renormalization of the effective mass of electrons or the effect of more bands. To account for this discrepancy, we introduced a renormalization factor into the Fermi velocities consistent with a renormalization of the effective mass of $\approx 2.0$. We note that a small measuring error in the assessment of the superconducting critical temperature $T_c$ will have a strong impact on estimated the renormalization of the effective mass. For example if $T_c$ shifts to a higher temperature just by 0.1 K, the renormalization value will be strongly reduced. Furthermore, the broad normal-superconductor transition displayed in the magnetoresistivity curves does not allow a very accurate determination of its onset value.

We have to distinguish two distinct sets of simulations: the first for $H_{c2}$ displayed in Fig. 4 and the second of the magnetoresistivity displayed in Figs. 2 and 3. The samples used for measuring $H_{c2}$ and magnetoresistivity have different critical temperatures and consequently we have used different parameters for their simulations.

The parameters used to simulate $H_{c2}$ and IMR are summarized in the Supplementary Table 2. $T_c$ is defined by experiment and critical temperatures $T'_{c1}$, $T'_{c2}$, $T'_{c3}$ can be obtained from $\bar{\lambda}_1, \bar{\lambda}_2, \bar{\lambda}_3, \gamma_{12}, \gamma_{23}$ and $\gamma_{13}$ by the expression $T'_{cl} = e^{C+2}\Omega_{BCS}e^{-1/\lambda_l}/\pi$, where $\lambda_l$ is related with $\bar{\lambda}_l$ and $\gamma_{ij}$ (see in text above) and C is the Euler constant. We expressed all coupling parameters as a function of a single coupling constant $\lambda$ and $T'_{cl}$ such that the superconducting critical temperature obtained from experiments is related with $\lambda$ by $T_c = e^{C+2}\Omega_{BCS}e^{-1/\lambda}/\pi$. The absolute values of the coupling constants $\bar{\lambda}_l$ for our compound were not yet determined by experiments. However, according to theoretical study in Ref. [9] the value of the strongest coupling constant for a similar doping levels is $\approx 2$, consequently we took $\lambda = 2$.

The set of parameters, $\bar{\lambda}_1, \bar{\lambda}_2, \bar{\lambda}_3, \bar{\gamma}_{12}, \bar{\gamma}_{23}$ and $\bar{\gamma}_{13}$ is not unique since, the values of the coupling constants, $\bar{\gamma}_{12}, \bar{\gamma}_{23}$ and $\bar{\gamma}_{13}$, between different components can be changed arbitrarily such that simulation results of $H_{c2}$ and IMR remain almost unchanged, if we also change the critical temperatures or the parameters $\bar{\lambda}_i$ associated with $\bar{\phi}_i$, where $i \in \{1,2,3\}$. Additionally, the highest $\bar{\lambda}_i$, designated by $\bar{\lambda}_m$ is defined such that $T'_{cm} = T_c$.

Supplementary Figs. 19 and 20 display the simulation results corresponding to the parameters in and fifth and seventh rows and in the first and fourth rows of Supplementary Table 2, respectively. The simulations corresponding to second, third, and sixth rows of the same table are displayed in Figs. 3 and 2.

In the Fermi surface there are electron pockets each on sides of the Brillouin zone, see Supplementary Figure 18. They are eccentric which allows the mixing between harmonic functions which have $s_\pm$-wave and $d_{x^2-y^2}$-wave symmetries in the centers of the pockets, displayed in Supplementary Eqs. (9)-(14), Eqs. (40)-(42) and Eqs. (43)-(45). Note that both electron pockets will compensate each other and will have well defined fourfold symmetry. To take into account the mixing for each component of the order parameter, we set a fix mixing of 10% between the harmonic functions with the same symmetry of the component and the harmonic functions which are allowed to be mixed on the electron pockets, i.e. $a_{i,2}/a_{i,1} = 0.1$ where $i \in \{1,2,3\}$. The absolute values of $a_{i,1}$ and $a_{i,2}$ can be found by the normalization of $\bar{\phi}_i$. We varied this ratio value and found that it only influences very weakly the $H_{c2}$ and magnetoresistivity results.

In the calculation of magnetoresistivity $U_0 = A\Delta G \frac{\Phi_0}{B} \xi(T)$ where A is a proportionality factor that was extracted by fitting the simulated to the experimental curves, $A = 0.01$. The value of $\xi(T) = \xi_0(1 - T/T_c)$ was set to $\xi_0 = 3.53$ nm according to Ref. [2].

## Supplementary Note 7. Estimation of the impact of misalignment in $H_{c2}$

We can assess the impact of misalignment in $H_{c2}$ by calculating the relative deviation of the magnetic critical field, i.e.

$$d(\varphi) = \frac{B_{c2}(\varphi,0) - B_{c2}(0,0)}{B_{c2}(0,0)}. \tag{62}$$

For this calculation we will consider only a single $s$-wave component since magnetic field cannot the mix the $s$-wave and the $d$-wave components. We will present here mathematical expressions that are shown, below, in theoretical description. We can obtain $B_{c2}(\varphi, \theta)$ using Supplementary Eq. (44).

If we only take into account the $\phi_1$ component, it becomes:

$$0 = w_{(0,0),1} - B_{c2} w_{(2,0),1}(\phi_1^2). \tag{63}$$

In the case of a $s$-wave component $\phi_1 = 1$. For simplicity we will drop the index 1 in all the coefficients and functions, and explicitly express the dependency of the variables on the angles, $\varphi$ and $\theta$, that define the direction of the applied field. The previous equation becomes

$$0 = w_{(0,0),1} - B_{c2} w_{(2,0),1}(\varphi, \theta). \tag{64}$$

Thus,

$$B_{c2} = \frac{w_{(0,0),1}}{w_{(2,0),1}(\varphi, \theta)}, \tag{65}$$

where

$$w_{(2,0),1}(\varphi, \theta) = \langle |\bar{v}_F(\varphi, \theta)|^2 \rangle \sum_{n=0}^{\infty} \frac{(2\pi)^2 T \hbar^2}{2^3 \varepsilon_n^3 \Phi_0}. \tag{66}$$

The relative deviation of $B_{c2}$ becomes:

$$d(\varphi) = \frac{B_{c2}(\varphi, 0) - B_{c2}(0,0)}{B_{c2}(0,0)} = \frac{\langle |\bar{v}_F(0,0)|^2 \rangle}{\langle |\bar{v}_F(\varphi, 0)|^2 \rangle} - 1. \tag{67}$$

We can find the value of $\bar{v}_F$ using Supplementary Eq. (37), where $v_{Fx}$ and $v_{Fy}$ are given by Supplementary Eqs. (40) and (41) respectively and $c_1$ and $c_2$ are given by:

$$c_1 = \left( \frac{\langle v_{Fy}^2 \rangle^2}{\langle v_{Fx}^2 \rangle \langle v_{Fy}^2 \rangle - \langle v_{Fx}^2 v_{Fy}^2 \rangle} \right)^{1/4}, \tag{68}$$

$$c_2 = \left( \frac{\langle v_{Fx}^2 \rangle^2}{\langle v_{Fx}^2 \rangle \langle v_{Fy}^2 \rangle - \langle v_{Fx}^2 v_{Fy}^2 \rangle} \right)^{1/4} \exp\left( i \tan^{-1}\left( \frac{-\langle v_{Fx} v_{Fy} \rangle}{\sqrt{\langle v_{Fx}^2 \rangle \langle v_{Fy}^2 \rangle - \langle v_{Fx}^2 v_{Fy}^2 \rangle}} \right) \right) \tag{69}$$

Using the Fermi velocities and the Fermi surface of the tight-binding model we estimate that a misalignment of 2 degrees induces a $d(2°)=0.033\%$, which is negligible when compared with the strong anisotropy of 14.4% in $H_{c2}$.

## Supplementary Note 8. Metastable states in measurements of the angle-dependent in-plane magnetoresistivity

The procedure taken for measuring angular dependence of IMR (displayed in all figures to the exception of Supplementary Figure 20) consists in lowering the temperature, with sample is along *a*/*b*-axis (we defined it as $\theta = 0°$), until the temperature at which we want to measure to IMR. Then, we rotate the sample along the *c*-axis and measured the IMR for each angle (with a given fixed angle step). While lowering the temperature along the *a*/*b*-axis, we can be trapping the system into a state that is a stable minimum for that direction but not for the other directions for which the IMR is going to be measured when the sample is rotated. To assess the metastability of the system we measured IMR for each desired angle after rotating the sample at 150 K and then lowering temperature until a given value. Supplementary Figure 18 show values of IMR taken using this new methodology and contrasts then with IMR values taken as all the other measurements in this work (as described in the beginning of the section). In this figure, we can observe that both methodologies give similar results, thus, excluding the presence of metastable states that explain the $C_2$ shape in the angular dependence of the IMR.

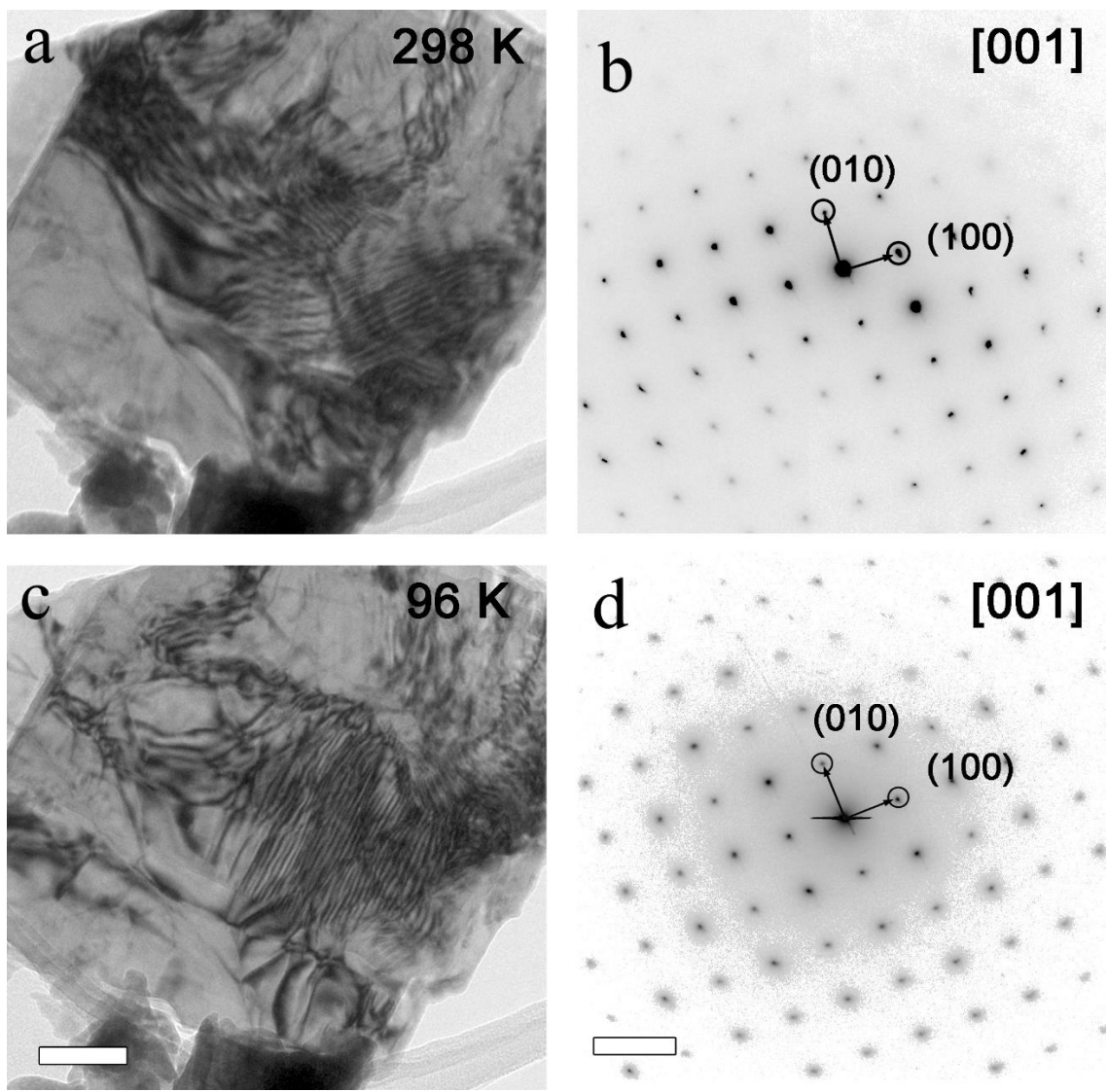

**Supplementary Figure 1. Low temperature TEM measurements.** (a) TEM image of a crystal in temperature of 298 K and (c) 96 K taken from the [001] direction. The corresponding SAED patterns at the marked positions on the crystal in (b) 298 K and (d) 96 K. Low temperature TEM measurements show the absence of intrinsic twin boundaries at temperature above 96 K, which is consistent with the low temperature XRD analysis on the optimally doped $(Ba,K)Fe_2As_2$ crystals [1]. Scale bar in (c) and (d) are 100 nm and 5 /nm, respectively.

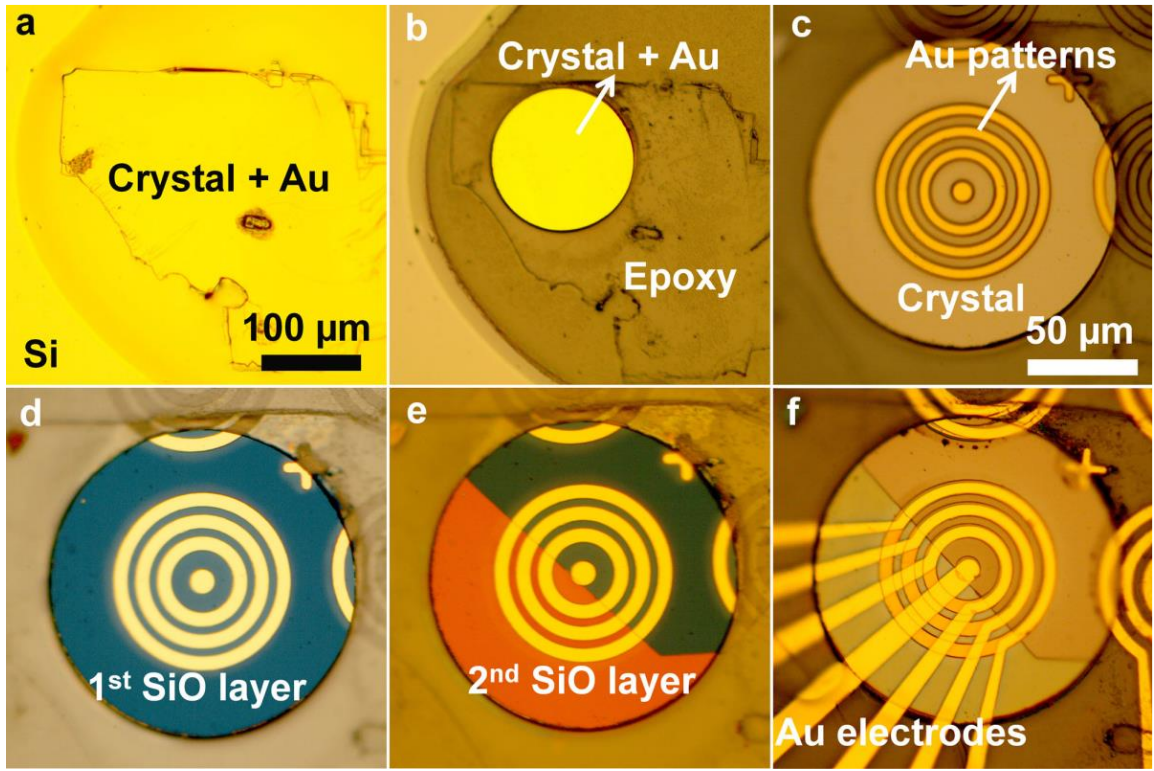

**Supplementary Figure 2. Optical photos for the Corbino sample in different processes.** Diagram of fabrication process for the Corbino disk device. (a) A cleaved crystal was coated with an approximately 120 nm layer of Au. (b) The flake crystal was fabricated into a disk shape. (c) A set of concentric circles of Au patterns was etched on the *ab*-plane. (d) A 100-nm insulating SiO layer was coated to surround the edges of the mesa. (e) Half of the mesa was covered with a 100-nm insulating SiO layer. (f) The electrodes were formed on the circular mesa by photolithography and argon-ion etching.

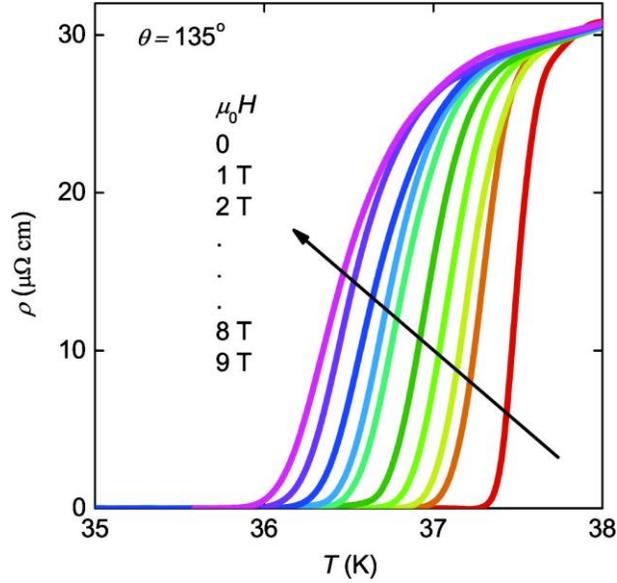

**Supplementary Figure 3. Temperature dependence of magnetoresistivity.** Temperature dependence of magnetoresistivity for the optimal-doped single crystal $Ba_{0.5}K_{0.5}Fe_2As_2$ under magnetic fields from 0 to 9 T. Here the magnetic fields were applied within the *ab*-plane with angle $\theta = 135°$. A superconducting transition shows an onset temperature of 37.6 K and a zero-*R* temperature of 37.3 K without a magnetic field. This sharp transition suggests that the quality of the single crystal is high, although the transition became wider with increasing magnetic field. The angle $\theta$ is defined as that between the magnetic field and the *a(b)*-axis of the lattice, as indicated in Fig. 1a. The dependence of $\rho_{ab}$ on $\theta$ was measured by rotating the *ab*-plane around the *c*-axis in a fixed magnetic field parallel to the *ab*-plane and at a fixed temperature.

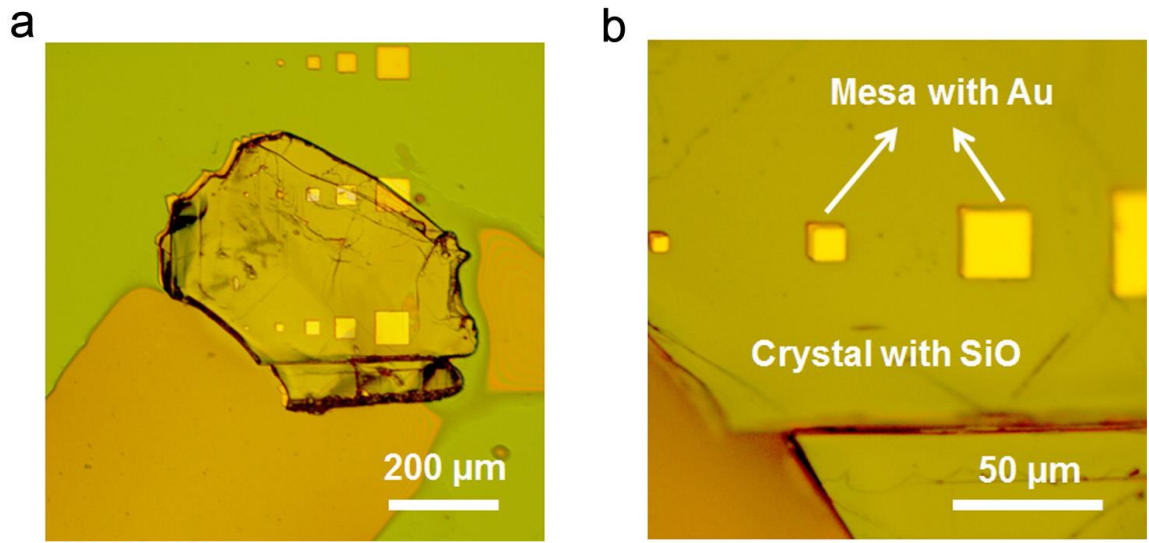

**Supplementary Figure 4. Optical photos for mesa sample.** Diagram of fabrication process for the mesa device. (a) Mesas were fabricated on a cleaved crystal. (b) Enlarged view of the mesas (10×10, 20×20, and 40×40 μm$^2$ in area and 1.5 μm in thickness from left to right). The thickness of the base crystal was larger than 20 μm, and the in-plane geometry was considerably larger as few hundred micrometers. Therefore, the resistance of the base crystal was extremely lower than that of the mesa.

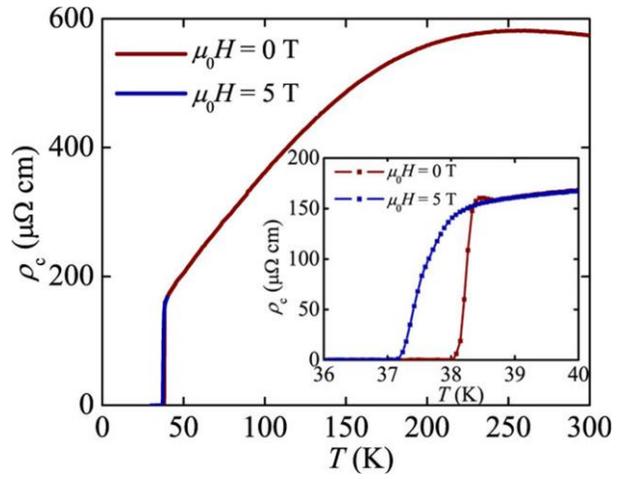

**Supplementary Figure 5.    Out-of-plane resistivity.** Temperature dependence of out-of-plane resistivity $\rho_c$ for Ba$_{0.5}$K$_{0.5}$Fe$_2$As$_2$ in a magnetic field of 5 T. The angle between the field and $a(b)$-axis was 45°.

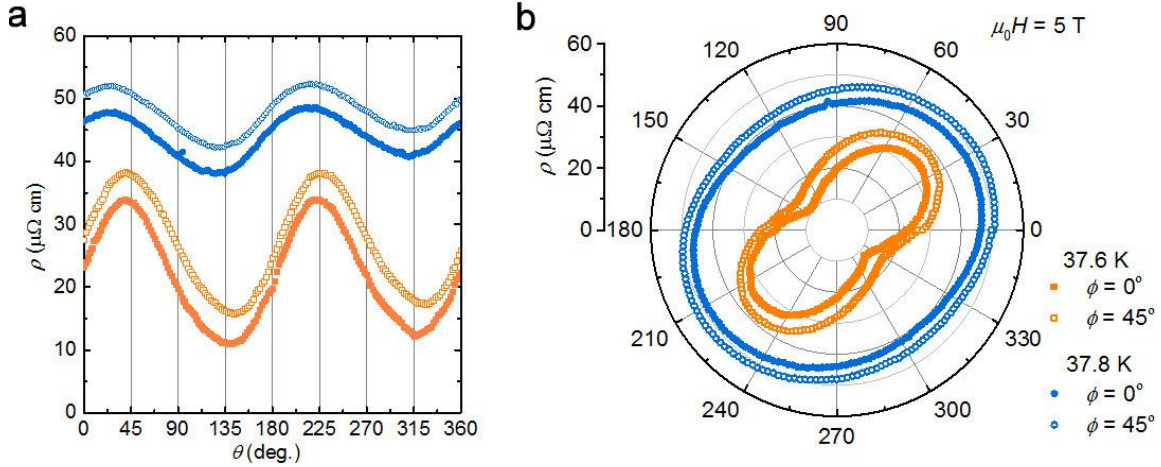

**Supplementary Figure 6. Out-of-plane magnetoresistivity under in-plane fields.** Angular dependence of out-of-plane magnetoresistivity for $Ba_{0.5}K_{0.5}Fe_2As_2$ in an in-plane magnetic field of 5 T for (a) and the corresponding polar plots for (b). The $\rho_c$ measurements were conducted at initial angles ($\phi$) of 0 and 45° between the field and $a(b)$-axis. As expected, the $\rho_c$-$T$ curve is slightly different from the $\rho_{ab}$–$T$ curve. The anisotropy parameter is 2.3 at 300 K, suggesting a quite weak electronic dimensionality of two dimensions. The angle-dependent out-of-plane magnetoresistivity shows an anomaly similar to that observed in the IMR measurements. Supplementary Figure 6 shows the out-of-plane magnetoresistivity at 37.6 and 37.8 K under a magnetic field of 5 T. The $\rho_c$-$\theta$ curves are observed to undergo sinusoidal oscillation with the maximum at the angle where $H$ is parallel to the ΓM direction, comparable to the IMR results. In addition, we tested the dependence of $\theta$ on the initial angle. Again, we confirmed that the angular dependence of the magnetoresistivity within the $ab$-plane truly reflects the in-plane nature.

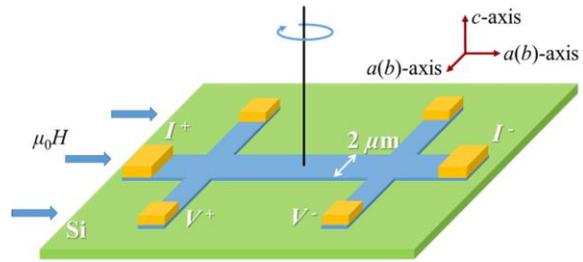

**Supplementary Figure 7.   Schematic image of a microbridge for pulsed high magnetic experiments.** The sample was rotated along the *c*-axis, and the magnetic field was applied within the *ab*-plane.

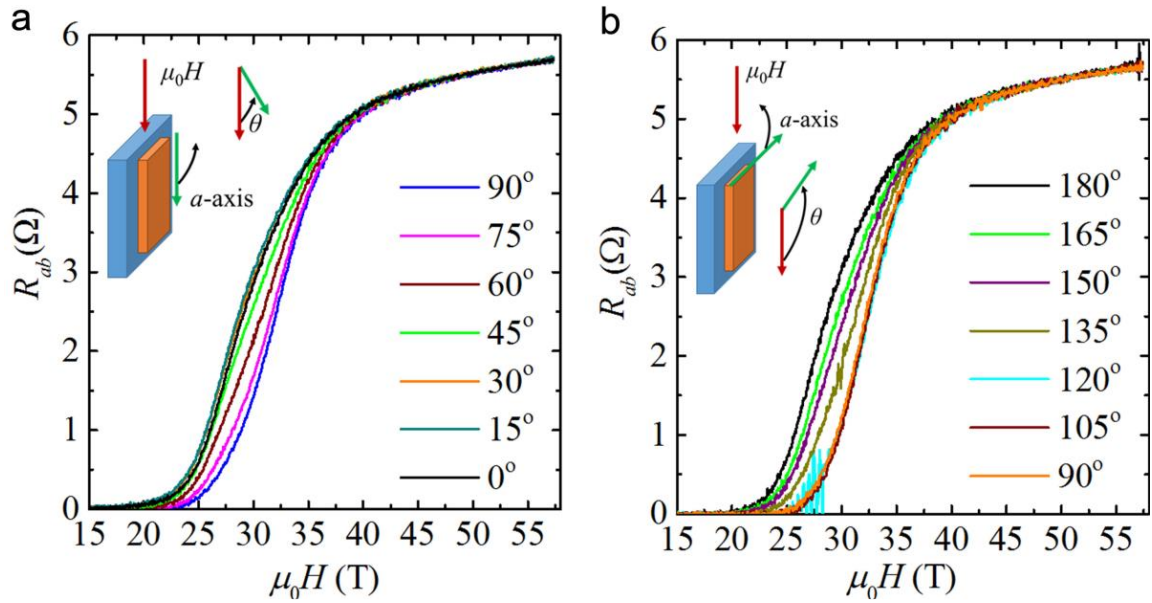

**Supplementary Figure 8. In-plane magnetoresistance for Ba$_{0.5}$K$_{0.5}$Fe$_2$As$_2$ under pulsed high magnetic fields up to 57 T.** Since the rotation system can provide only 90º rotating angle, we measured the sample within two steps, namely, from 0 to 90º in (a) and 90º-180º in (b), respectively. Here the temperature was fixed at 35 K, and the $\theta$ was defined as the angle between the magnetic field and the *a*-axis.

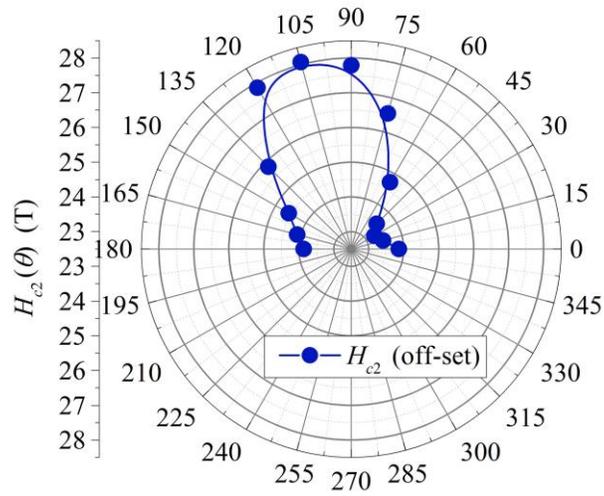

**Supplementary Figure 9.  Angular dependence of the second magnetic upper critical field.** Here the $H_{c2}(\theta)$ data were estimated from the pulsed high magnetic field experiments in Supplementary Figure 9, and are consistent with the static magnetic fields experiments.

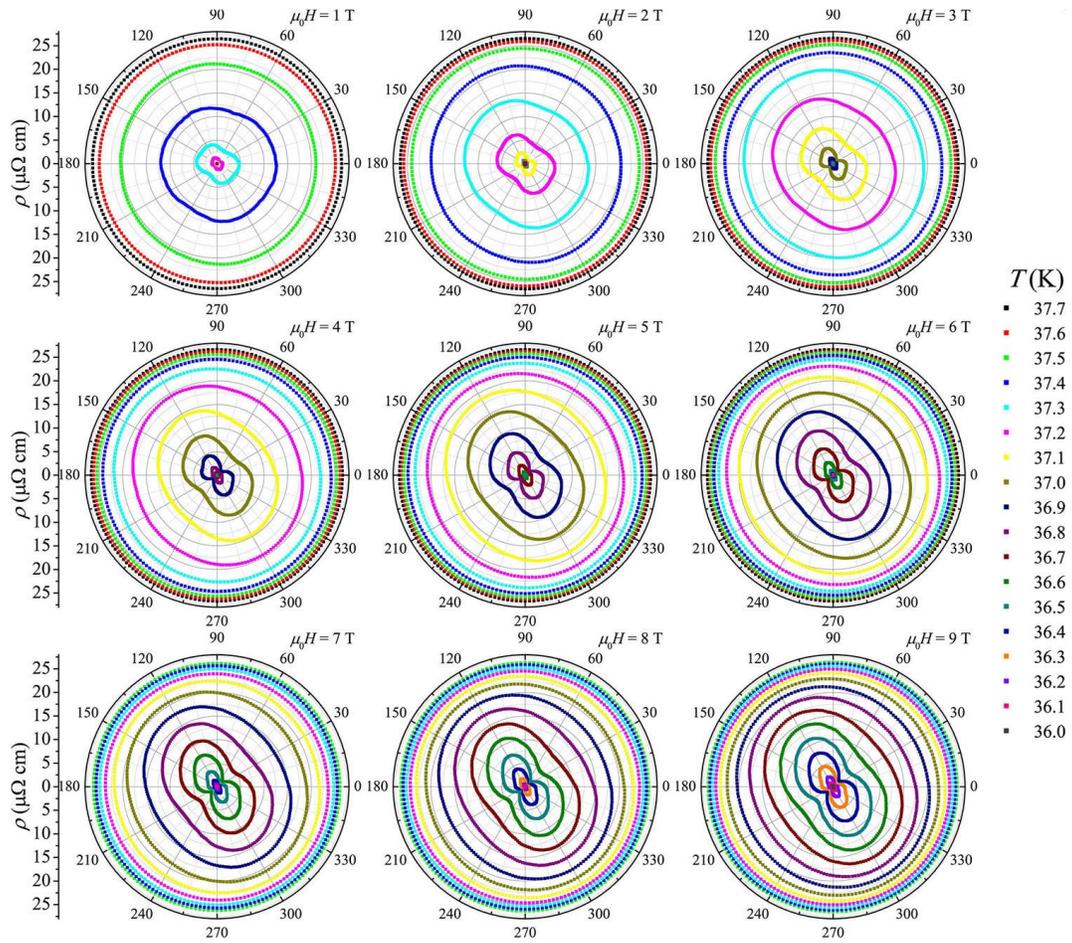

**Supplementary Figure 10.    Polar plots of in-plane magnetoresistivity.** Here, the data are from Fig. 2 for the $Ba_{0.5}K_{0.5}Fe_2As_2$ Corbino sample at various temperatures and magnetic fields. Data for weaker fields show a weakly fourfold symmetric anomaly. The fourfold anomaly gradually transforms to a strong two-fold symmetric anomaly as the magnitude of the applied magnetic field increases.

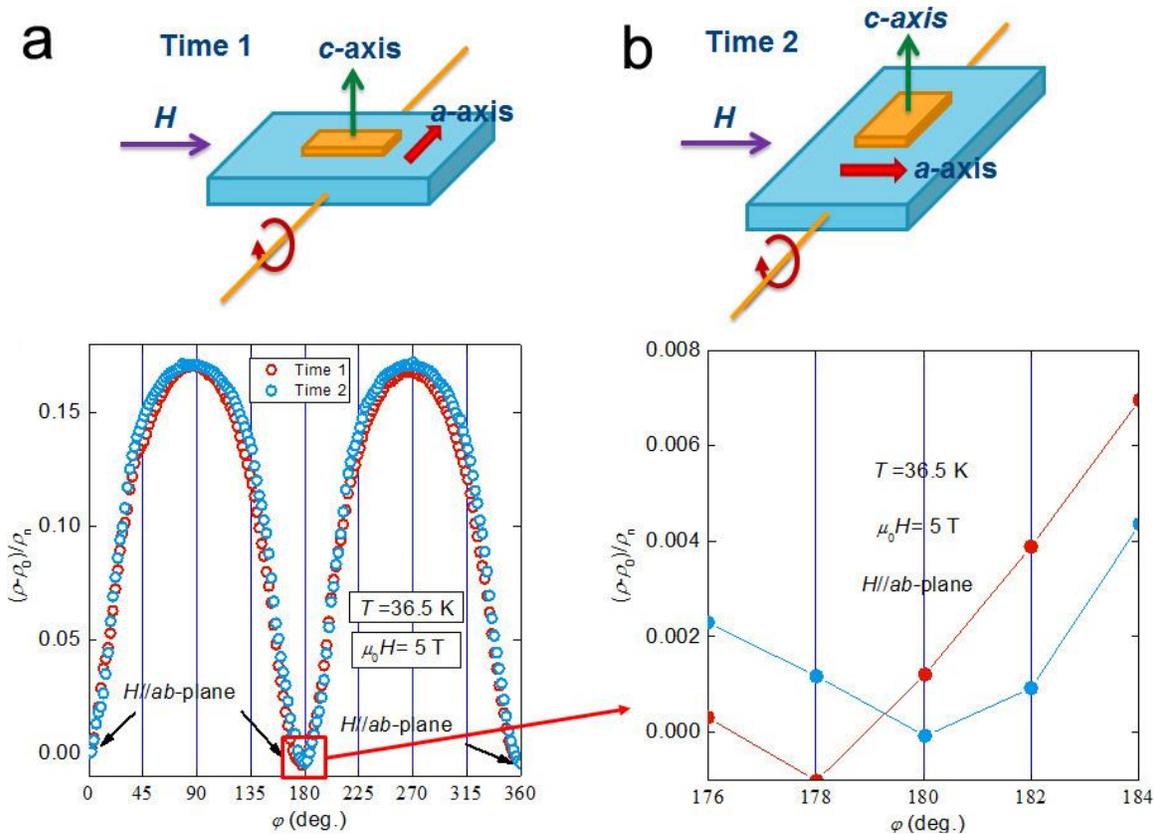

**Supplementary Figure 11. Out-of-plane magnetoresistivity measurements.** (a) Angular dependence of out-of-plane magnetoresistivity for $Ba_{0.5}K_{0.5}Fe_2As_2$. The magnetic field was rotated from the $a(b)$-axis to the $c$-axis. $\varphi$ is the angle between the $a(b)$-axis and $H$. (b) The enlarged view for the field parallel with the $ab$-plane. Up figures are the schematic image for measurement setup. These measurements is to evaluate a possible misalignment between the crystal $c$-axis and the mechanical rotating axis of the sample stage. The maxima appear at exactly 90° and 270° as expected, and the magnetic field is precisely perpendicular to the $ab$-plane, indicating that the $ab$-plane is adequately parallel to the substrate and the sample stage. From the out-of-plane magnetoresistivity data, we estimated that the misalignment is less than ±1°. Even if the angle error is up to ±5°, the influence on the normalized IMR is still very small (e.g., <0.002 at 37.0 K under $\mu_0 H = 5$ T, compared to the normalized IMR of −0.20 to 0.15).

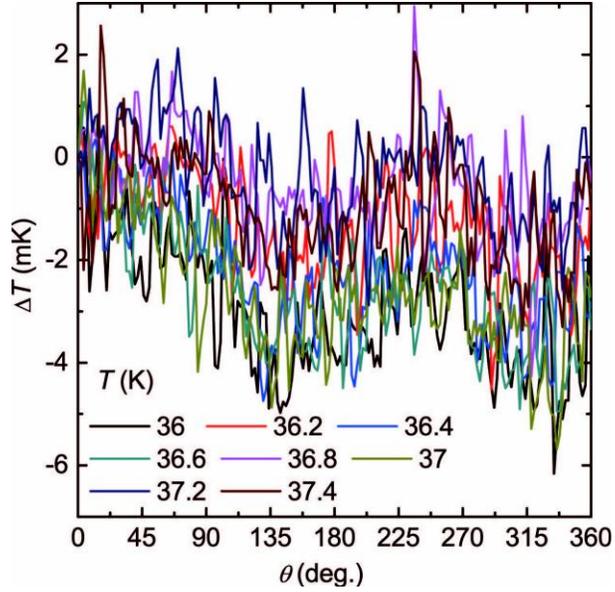

**Supplementary Figure 12. Temperature fluctuation.** Angular dependence of normalized sample temperature $\Delta T = T - T_{\text{bath}}$, where, $T$ means the temperature of the sample at different anglers, and $T_{\text{bath}}$ the bath temperature. The applied field $\mu_0 H$ was 9 T. This experiment is to evaluate the temperature variation of the sample stage in the PPMS, where the angular dependence of temperature is measured at zero-field. The result shows a very weak variation of the sample stage temperature regardless of the angle. The sample temperature difference during rotation is less than 0.008 K. As a result, we can hardly ascribe the large anisotropic IMR data to a temperature problem.

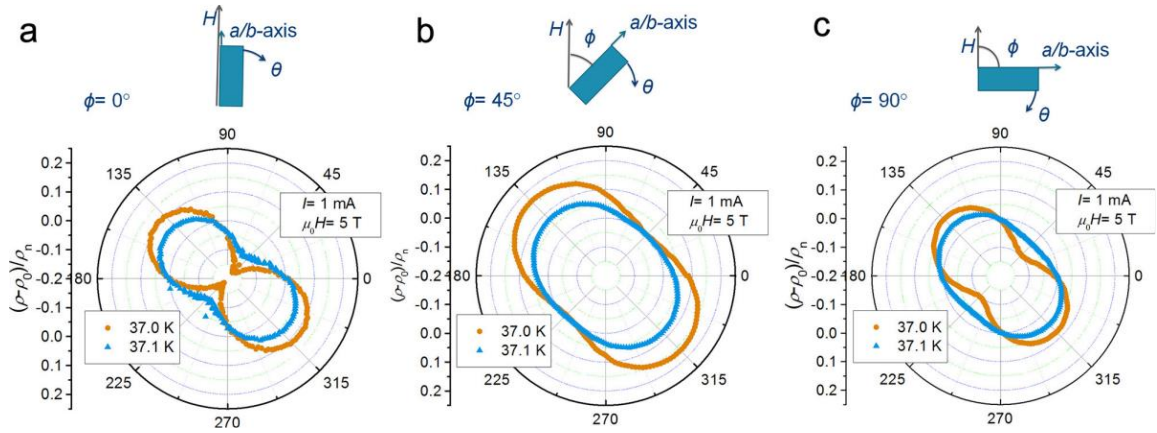

**Supplementary Figure 13. Initial anger dependent measurements.** Polar plots of normalized IMR with initial angles ($\phi$) of (a) 0°, (b) 45°, and (c) 90°. The applied field was $\mu_0 H = 5$ T, and the sample temperatures were 37.0 and 37.1 K. The polar angle $\theta$ is an angle between the magnetic field and $a(b)$-axis. The maximum (minimum) $\rho_{ab}(\theta)$ was observed along the Fe-Fe direction as well.

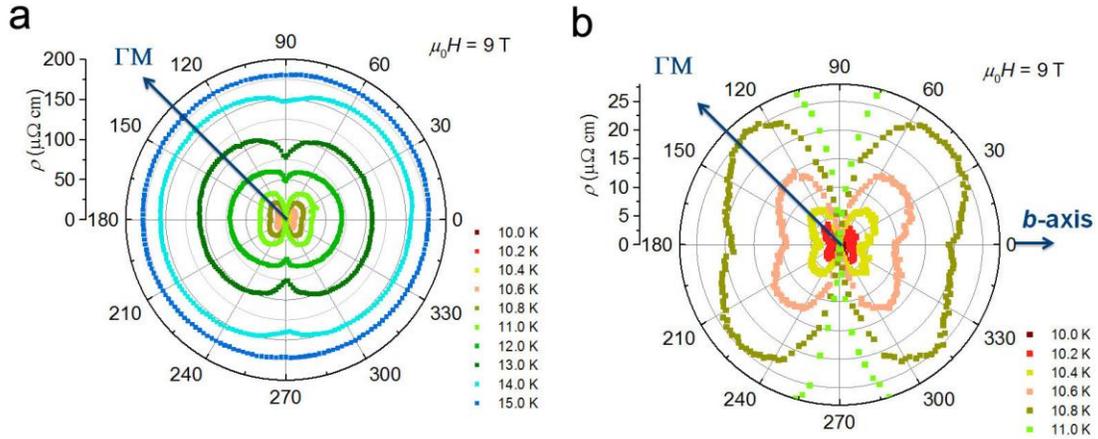

**Supplementary Figure 14. Under-doped measurements.** (a) Angular dependence of IMR for under-doped single crystal $Ba_{0.75}K_{0.25}Fe_2As_2$ in a magnetic field of 9 T. Here, $\theta$ is an angle between the magnetic field and *b*-axis. (b) Angular dependence of the IMR around $T_c$-onset, where the symmetry can be observed as a mixture four-fold and two-fold with minimum corresponding to *a*- and *b*-axis. Note that the *a*-axis minimum is lower than that of *b*-axis, which is probably owing to the two-fold structural distortion induced nematic state [12]. The tetragonal phase of the compound family is of the $ThCr_2Si_2$-type structure (space group *I*4/*mmm*) and the orthorhombic phase is of the *β*-$SrRh_2As_2$-type structure (*F/mmm*), see Ref. [1]. We note that the *a*- and *b*-axis of the tetragonal lattice structure are rotated 45 degrees from the orthorhombic lattice structure.

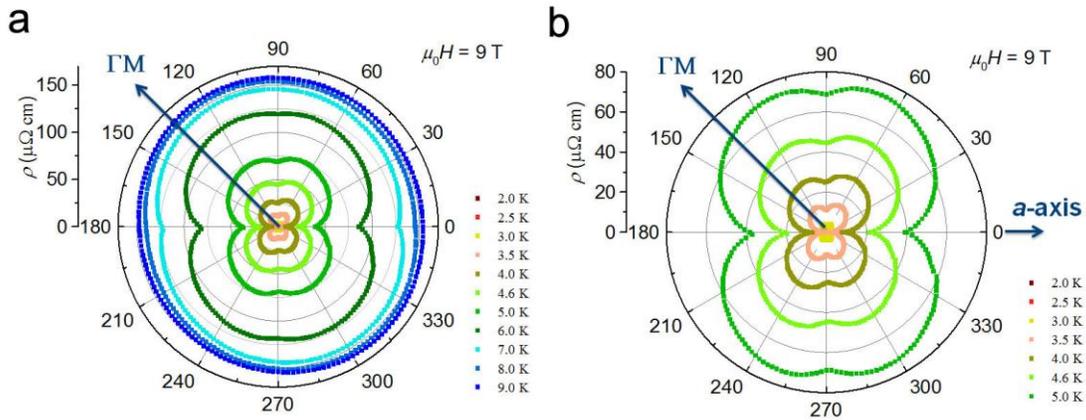

**Supplementary Figure 15**. **Under-doped measurements for another sample.** (a) Angular dependence of IMR for under-doped single crystal $Ba_{0.8}K_{0.2}Fe_2As_2$ in a magnetic field of 9 T, and (b) the enlarged view of IMR around $T_c$-onset. Here, $\theta$ is an angle between the magnetic field and *a*-axis. The symmetry can be also observed as a mixture four-fold and two-fold as those of $Ba_{0.75}K_{0.25}Fe_2As_2$ sample. The anisotropic behavior between the *a*-axis and *b*-axis may be due to the structural distortion induced nematic state as well.

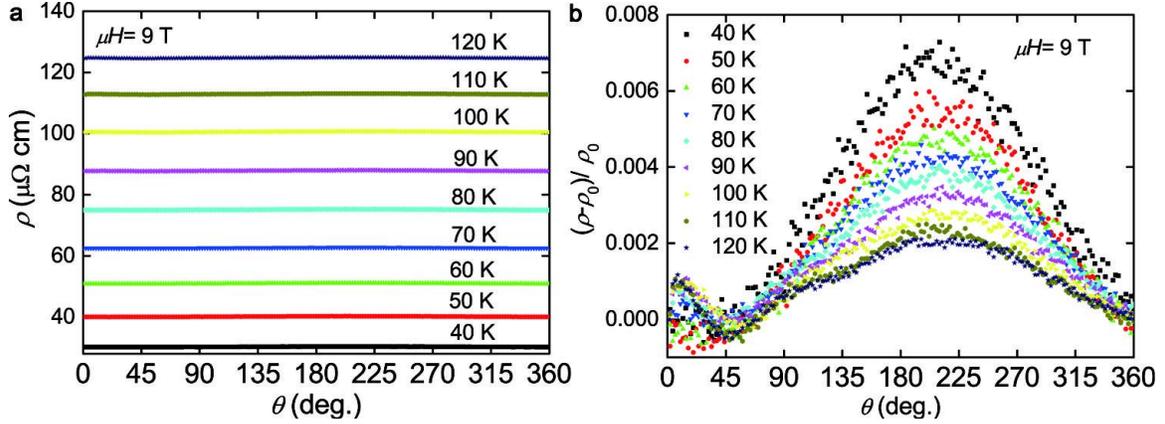

**Supplementary Figure 16. Normal state measurements. Angular** dependence of IMR (a) and normalized IMR (b) in the normal phase for the single crystal $Ba_{0.5}K_{0.5}Fe_2As_2$ in a magnetic field of 9 T, from 40 K to 120 K. The dominant $C_1$ feature in the angular dependence (b) cannot be attributed either to resistivity or magnetoresistivity angular oscillation because any of them should display invariance under space inversion. The one-fold symmetry of the angular dependence can only be due to drawbacks in the measuring procedure. As a result, the amplitude of the two-fold symmetry oscillation of (magneto) resistivity, signaling small tetragonal symmetry breaking in the sample due to anisotropy of internal stresses, is expected to be lower than the total amplitude of oscillation displayed in plot (b).

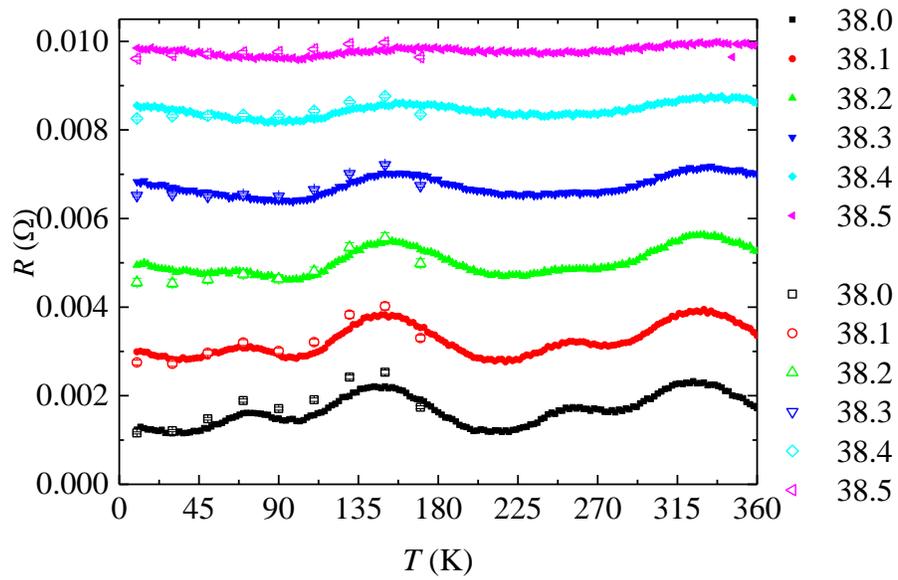

**Supplementary Figure 17. IMR measurements after high temperature (150 K) treatment.** Angular dependence of IMR, from 38.0 K to 38.5 K, in resistance domain of the normal to superconductor transition after the nucleation of superconductivity. Filled and unfilled symbols correspond to values of IMR measured after the rotation of the sample at the measured temperature or at 150 K temperature.

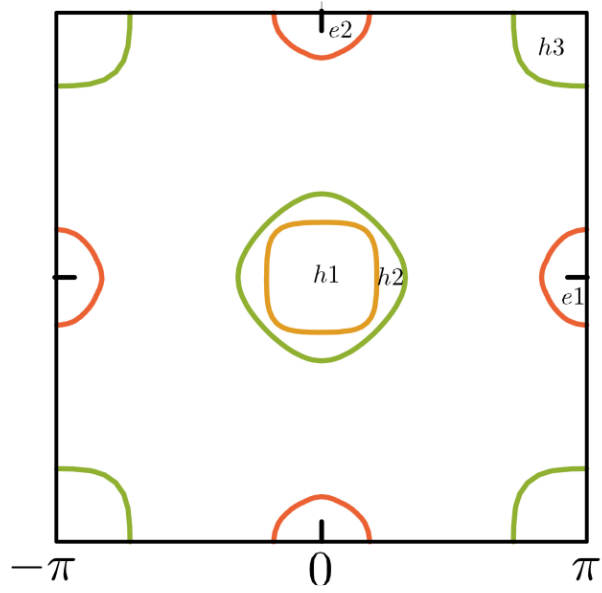

**Supplementary Figure 18. Cut of the Fermi surface of $Ba_{0.5}K_{0.5}Fe_2As_2$.** The Fermi surface is in the $Z = 0$ plane of the extended Brillouin zone (*X*- and *Y*-axis of the conventional lattice are aligned along the Fe-Fe bond directions) corresponding to the **Z**-axis direction of the conventional lattice. Lines in green, yellow and red represent the Fermi surface in distinct bands. The Fermi surface is disconnected and can be subdivided into three hole pockets, *h*1, *h*2 and *h*3 and two electron pockets *e*1 and *e*2, identified inside each pocket the figure.

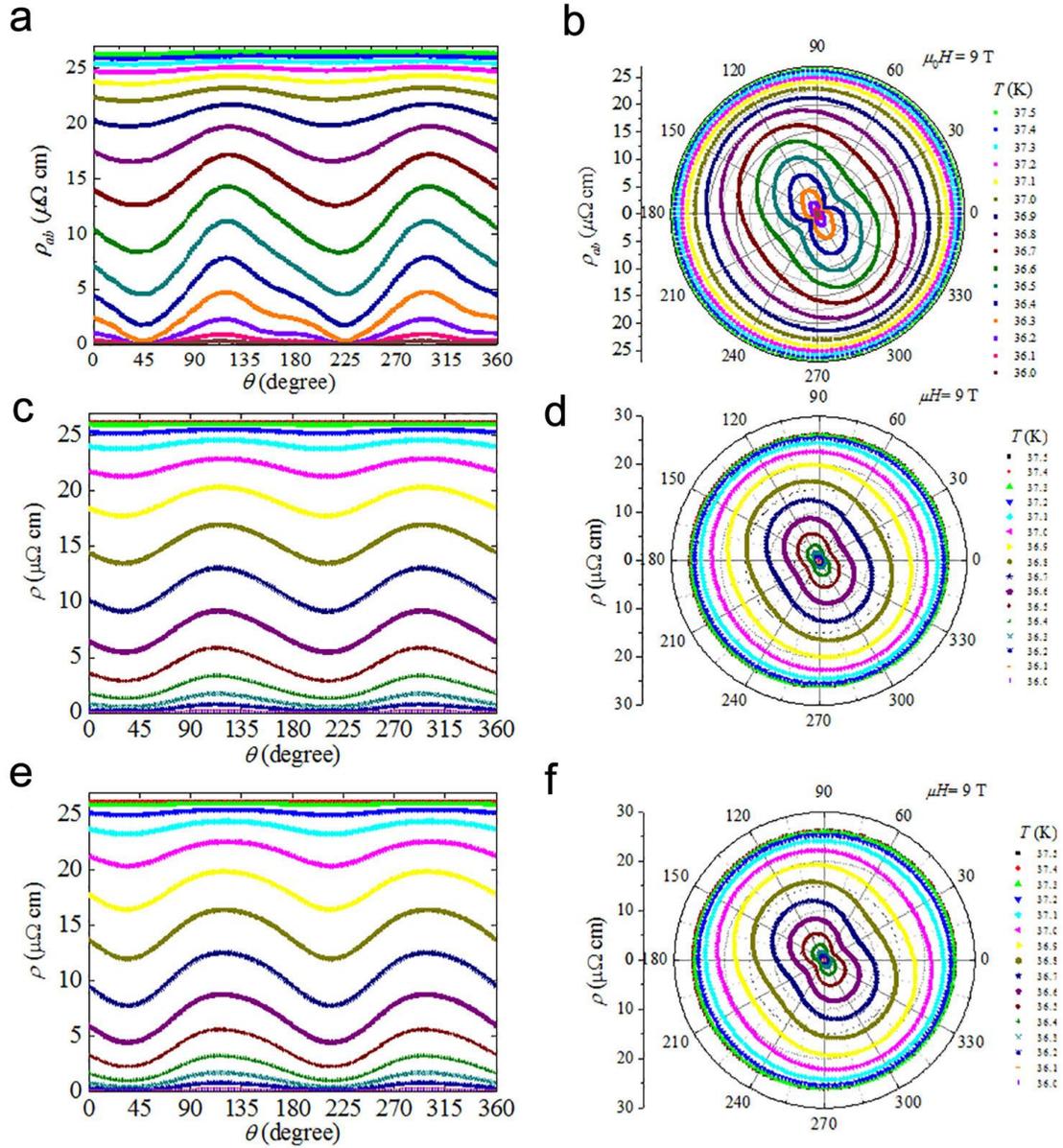

**Supplementary Figure 19. Experimental and theoretical of the angular dependence of the IMR.** Experimental (a) and theoretical (c, e) values of the angular dependence of the IMR, and respective polar plots of IMR experimental (b) and theoretical (d, f) values, at various temperatures for the applied magnetic field of 9 T for which the experimental values were obtained using the Corbino disc measurement configuration. Theoretical values (c, e) of the angular dependence of the IMR, refer to the models where the order parameter has components $s_\pm$-wave, $d_{x^2-y^2}$-wave symmetry and an additional $d_{xy}$-wave and $g$-wave symmetry component, respectively.

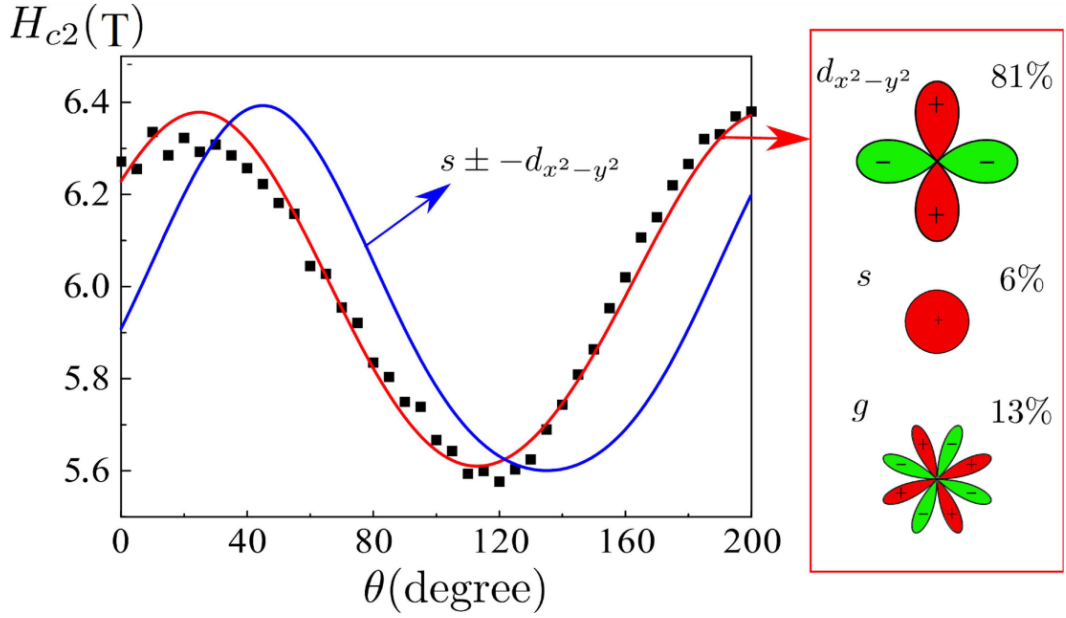

**Supplementary Figure 20. Angular dependence of the second magnetic critical field at 38.4 K ($T_c \approx 39$ K)**. Here, the data are retrieved from transport experiments (black filled square symbol) in the Corbino disc measurement configuration, from theoretical model with *s*-wave and $d_{x^2-y^2}$-wave symmetries (full blue line) and from theoretical model with $s_\pm$-wave, $d_{x^2-y^2}$-wave and *g*-wave symmetries (full red line). The full lines in green and blue correspond to parameter values in the first and fourth rows of Supplementary Table 2. The mixing of the different symmetry components of the order parameters is indicated on the right side panel next to a schematic representation of each component of the order parameter as function of the internal momentum of the Cooper pairs. The indicated percentages correspond to the relative weights ($r_1$, $r_2$ and $r_3$) of the wave function coefficients, $\boldsymbol{\phi}(\mathbf{k}) = r_1 \boldsymbol{\phi}_s(\mathbf{k}) + r_2 \boldsymbol{\phi}_{d_{x^2-y^2}}(\mathbf{k}) + r_3 \boldsymbol{\phi}_{d_{xy}}(\mathbf{k})$. In this schematic red and green indicate positive and negative value of the components, respectively.

**Supplementary Table 1:** Deviations of the IMR values presented in Supplementary Figure 19 from the average value, i.e. **Dev.** $= 100 \frac{\text{Max}(\rho(\theta)) - \text{Min}(\rho(\theta))}{\overline{\rho(\theta)}}$. Shown symbols correspond to curves in the figure with the same symbols.

| Symb. | ■ | ● | ▲ | ▼ | ◆ | ◀ | ▶ | ⬢ | ★ |
|---|---|---|---|---|---|---|---|---|---|
| $T$ | 40 K | 50 K | 60 K | 70 K | 80 K | 90 K | 100 K | 110 K | 120 K |
| Dev. | 0.79% | 0.68% | 0.56% | 0.48% | 0.41% | 0.35% | 0.31% | 0.31% | 0.26% |
| Max. | 30.40 | 40.33 | 51.06 | 62.82 | 75.25 | 87.99 | 100.68 | 113.07 | 124.86 |
| Min. | 30.16 | 40.06 | 50.78 | 62.52 | 74.94 | 87.68 | 100.37 | 100.37 | 124.54 |

**Supplementary Table 2:** Parameters associated with simulations of $H_{c2}$ and in-plane magnetoresistivity, IMR, (first column-"Simul.") for models with two components, with $s\pm$- and $d_{x^2-y^2}$-wave symmetry, and optionally a third component, which can be of $g$- or $d_{xy}$-wave symmetry (second column-"3rd").

| Simul. | 3rd | $T_c$ | $T'_{c1}$ | $T'_{c2}$ | $T'_{c3}$ | $\bar{\lambda}_1$ | $\bar{\lambda}_2$ | $\bar{\lambda}_3$ | $\bar{\gamma}_{12}$ | $\bar{\gamma}_{23}$ | $\bar{\gamma}_{13}$ |
|---|---|---|---|---|---|---|---|---|---|---|---|
| $H_{c2}$ | $g$ | 39 | 29.92 | 39 | 38.22 | 1.3161 | 1.9928 | 1.9204 | 0.0607 | -0.0159 | 0.0467 |
| $H_{c2}$ | $d_{xy}$ | 39 | 28.78 | 39 | 37.81 | 1.2543 | 1.9872 | 1.8853 | 0.0463 | -0.0283 | -0.0713 |
| $H_{c2}$ | $d_{xy}$ | 39 | 39 | 30.45 | 25.96 | 1.9907 | 1.3390 | 1.1110 | 0.0464 | -0.0284 | -0.0714 |
| $H_{c2}$ | none | 37.5 | 29.93 | 37.5 | - | 1.3159 | 1.9924 | - | 0.0723 | - | - |
| IMR | $g$ | 37.5 | 28.65 | 37.50 | 35.10 | 1.3177 | 1.9951 | 1.7533 | 0.0497 | -0.0227 | -0.0293 |
| IMR | $d_{xy}$ | 37.5 | 27.72 | 37.50 | 35.82 | 1.2543 | 1.9871 | 1.8371 | 0.0717 | -0.0293 | -0.0245 |
| IMR | $d_{xy}$ | 37.5 | 37.5 | 28.89 | 24.68 | 1.99 | 1.3143 | 1.0989 | 0.0495 | -0.0226 | -0.0777 |

**Supplementary References**